\documentclass[aps,prd,twocolumn,preprintnumbers,superscriptaddress,nofootinbib]{revtex4-1}
\usepackage{dcolumn}
\usepackage{amsmath}
\usepackage{amssymb}
\usepackage{amsthm}
\usepackage{mathtools}
\usepackage{mathrsfs}
\usepackage{bm}
\usepackage{slashed} 
\usepackage{graphicx}
\usepackage{multirow}
\usepackage{tikz}
\usepackage[caption=false]{subfig}
\usepackage{relsize}	
\usepackage{array}
\usepackage{float}
\usepackage{color}
\usepackage{xcolor}
\usepackage{soul}
\usepackage{verbatim} 
\usepackage{siunitx}
\usepackage{hyperref}
\hypersetup{colorlinks=true,linkcolor=purple,anchorcolor=blue,citecolor=purple1, filecolor=blue,urlcolor=red,bookmarksnumbered=true,
pdfview=FitB
}
\allowdisplaybreaks
\def\be{\begin{eqnarray}}
\def\ee{\end{eqnarray}}
\colorlet{purple1}{blue!70!red}
\colorlet{darkred}{red!50!black}

\begin{document}



\title{Generalized parton distributions and spin structures of light mesons from a light-front Hamiltonian approach}

\author{Lekha~Adhikari}
\email{adhikari@iastate.edu}
\affiliation{Department of Physics and Astronomy, Iowa State University,
Ames, IA 50011, U.S.A.}

\author{Chandan~Mondal}
\email{mondal@impcas.ac.cn} 
\affiliation{Institute of Modern Physics, Chinese Academy of Sciences, Lanzhou 730000, China}
\affiliation{School of Nuclear Science and Technology, University of Chinese Academy of Sciences, Beijing 100049, China}
\affiliation{CAS Key Laboratory of High Precision Nuclear Spectroscopy, Institute of Modern Physics, Chinese Academy of Sciences, Lanzhou 730000, China}

\author{Sreeraj~Nair}
\email{sreeraj@impcas.ac.cn} 
\affiliation{Institute of Modern Physics, Chinese Academy of Sciences, Lanzhou 730000, China}
\affiliation{School of Nuclear Science and Technology, University of Chinese Academy of Sciences, Beijing 100049, China}
\affiliation{CAS Key Laboratory of High Precision Nuclear Spectroscopy, Institute of Modern Physics, Chinese Academy of Sciences, Lanzhou 730000, China}

\author{Siqi~Xu}
\email{xsq234@impcas.ac.cn} \affiliation{Institute of Modern Physics, Chinese Academy of Sciences, Lanzhou 730000, China}
\affiliation{School of Nuclear Science and Technology, University of Chinese Academy of Sciences, Beijing 100049, China}
\affiliation{CAS Key Laboratory of High Precision Nuclear Spectroscopy, Institute of Modern Physics, Chinese Academy of Sciences, Lanzhou 730000, China}

\author{Shaoyang~Jia}\email{syjia@anl.gov}
\affiliation{Department of Physics and Astronomy, Iowa State University, Ames, IA 50011, U.S.A.}
\affiliation{Physics Division, Argonne National Laboratory, Argonne, IL 60439, U.S.A.}

\author{Xingbo~Zhao}\email{xbzhao@impcas.ac.cn}
\affiliation{Institute of Modern Physics, Chinese Academy of Sciences, Lanzhou 730000, China}
\affiliation{School of Nuclear Science and Technology, University of Chinese Academy of Sciences, Beijing 100049, China}
\affiliation{CAS Key Laboratory of High Precision Nuclear Spectroscopy, Institute of Modern Physics, Chinese Academy of Sciences, Lanzhou 730000, China}

\author{James~P.~Vary}\email{jvary@iastate.edu}
\affiliation{Department of Physics and Astronomy, Iowa State University,
Ames, IA 50011, U.S.A.}

\collaboration{BLFQ Collaboration}
\date{\today}

\begin{abstract}
We present the generalized parton distributions (GPDs) for the valence quarks of the pion and the kaon in both momentum space and position space within the basis light-front quantization framework. These GPDs are obtained from the eigenvectors of a light-front effective Hamiltonian consisting of the holographic quantum chromodynamics (QCD) confinement potential, a complementary longitudinal confinement potential, and the color-singlet Nambu-Jona--Lasinio interactions for the valence quarks of mesons. We then calculate the generalized form factors of the pion and the kaon from the moments of these GPDs. Combining the tensor form factors with the electromagnetic form factors, we subsequently evaluate the impact parameter dependent probability density of transversely polarized quarks inside the pion and the kaon. The numerical results for the generalized form factors, corresponding charges, as well as those for the probability densities and the transverse shift of the polarized densities are consistent with lattice QCD simulations and with chiral quark models.

\end{abstract}
\maketitle
\section{Introduction}
The description of the nonperturbative structure of hadrons using generalized parton distributions (GPDs) is related to phenomenology, therefore attracting numerous dedicated experimental and theoretical efforts~\cite{Ji:PRL,Polyakov:1999gs,Diehl:2003ny,Belitsky:2005qn,Goeke:2001tz,Choi:2001fc,Kaur:2018ewq,deTeramond:2018ecg,Ji:2006ea,Fanelli:2016aqc,Chouika:2017rzs,Mezrag:2014jka,Broniowski:2007si,Zhang:2021mtn,Zhang:2021shm,Hwang:2007tb,Mondal:2017wbf,Chakrabarti:2014cwa,Adhikari:2016idg,Adhikari:2018umb,Meissner:2007rx,Alexandrou:2020zbe,Chen:2019lcm,Dupre:2016mai,Kriesten:2021sqc}.  These  GPDs are experimentally accessible through exclusive processes including deeply virtual Compton scattering (DVCS) and deeply virtual meson production (DVMP). The GPDs present an attractive testing ground for comparing theory with experiment since they encode a wealth of information about the spatial structure of the hadron as well as the partonic distribution of spin and orbital angular momenta.
Unlike the parton distribution functions (PDFs), which are solely functions of longitudinal momentum fraction ($x$) carried by the active parton, the GPDs are  functions of $x$, the skewness ($\zeta$) which represents the longitudinal momentum transfer,  and the square of total momentum transfer ($t$) to the hadrons.

The GPDs provide a picture that unites PDFs with form factors (FFs), where the former  describe the longitudinal momentum distribution of partons within a hadron while the latter characterize the spatial extent. One obtains the FFs, charge distributions, PDFs, etc. from the GPDs by marginalizing~\cite{guidal-ff, nikkhoo-ff, miller-charge}. Additionally, in the absence of the longitudinal momentum transfer ($\zeta=0$), the GPDs are converted to the impact parameter dependent parton distributions via Fourier transform with respect to the transverse momentum transfer. Unlike the GPDs themselves, the impact parameter dependent parton distribution is the probability density of partons at a given combination of the longitudinal momentum fraction and the transverse distance from the center of the hadron~\cite{Burkardt:2002hr,Burkardt:2000za,Ralston:2001xs,Broniowski:2003rp}. For different polarizations of the partons, spin densities can be expressed in terms of the polarized impact parameter dependent GPDs~\cite{Brommel:2007xd,Gockeler:2006zu,Pasquini:2007xz,Maji:2017ill,Diehl:2005jf}.

For many years, DVCS and DVMP data have been accumulated by J-PARC, Hall-A and Hall-B of JLab by the CLAS collaboration and by COMPASS at CERN \cite{gpd-exp, gpd-exp1, gpd-exp2, gpd-exp3, gpd-exp4, gpd-exp6, gpd-exp7}. Recently, JLab has also started a positron initiated DVCS experiment~\cite{Accardi:2020swt}, COMPASS at CERN will start to collect more DVCS data, while future Electron-Ion Colliders~\cite{AbdulKhalek:2021gbh,Anderle:2021wcy} are planned to explore the GPDs through DVCS. However, experimental extractions of the GPDs are not straightforward.  In particular, fitting  of DVCS data does not provide direct information about the GPDs but, instead, provides  some weighted integrals of the GPDs. Since nonperturbative QCD predictions are not yet possible from the first principles, model predictions of the GPDs are useful for constraining the GPDs and  data fitting in order to develop insights  into GPDs from  DVCS data. 

Among known hadrons, the pion plays a leading role for comparing theory with experiment. From the Drell-Yan process  with pion beams \cite{Drell:1970wh, Christenson:1970um}, we can access the partonic structure of the pion by colliding them with nuclear targets \cite{Peng:2014hta, Chang:2013opa, Reimer:2007iy, McGaughey:1999mq}. 
 Chiral symmetry is dynamically broken in QCD leading to generation of the Goldstone bosons (pions) having a small mass when compared to other
hadrons. On the one hand, the pions are salient in providing the force that binds the neutrons and the protons inside the nuclei and they also affect the properties of the isolated nucleons. Hence one can safely say that our understanding of visible (baryonic) matter  is incomplete without detailed knowledge of the structure and interactions  of the pion.  
On the other hand, the pseudoscalar kaons,  counterparts
of the pions with one strange valence quark, play a critical role in our understanding of Charge and Parity (CP) symmetry
violation~\cite{Woods:1988za,Barr:1993rx,Gibbons:1993zq}. In this paper, we investigate the partonic structure of the pions and the kaons in terms of their GPDs.
As background, we note that  different theoretical analyses have provided useful insights  regarding the pion GPDs, \emph{e.g}.\ Refs.~\cite{Broniowski:2003rp,Polyakov:1999gs,Frederico:2009fk,Kaur:2018ewq,Theussl:2002xp,Dalley:2003sz,Broniowski:2007si,Mezrag:2014jka,Fanelli:2016aqc,Kumano:2017lhr,Ma:2019agv,Zhang:2020ecj,Shi:2020pqe,Gutsche:2015,Gutsche:2013zia,deTeramond:2018ecg,Chang:2020kjj,Brommel:2007xd,Hagler:2009ni,dalley,dalley1,Chen:2019lcm,Zhang:2021mtn,Roberts:2021nhw,Kaur:2020vkq,Raya:2021zrz}, while for the kaon, foundations are just being laid and several significant analyses can be found in Refs.~\cite{Kaur:2020vkq,Nam:2011yw, Xu:2018eii, Kock:2020frx,Kaur:2019jow,Zhang:2021mtn,Zhang:2021tnr,Raya:2021zrz}.

Another salient issue is the transversity of the hadrons~\cite{Barone:2001sp}, which provides access to their spin structures. Due to transversity's chiral-odd nature, it is challenging to measure  experimentally. Nevertheless, the transverse spin asymmetry in Drell-Yan
processes in $p\bar{p}$ reactions~\cite{Anselmino:2004ki,Pasquini:2006iv} and the azimuthal single spin asymmetry in semi-inclusive deep inelastic scattering (SIDIS)~\cite{Anselmino:2007fs} can be used to extract valuable  information on the transversity of the nucleon. While the transversity of the nucleon is nonzero and has now been well determined~\cite{Radici:2018iag}, it vanishes for the spin-zero hadrons. However, the chiral-odd GPDs defined as off-forward matrix elements of the tensor current are nonzero and much less information is available for them in the case of the pion and the kaon.

From the perspective of theory, the QCDSF/UKQCD Collaboration has reported the first result for the pion's chiral-odd GPD using lattice QCD~\cite{Brommel:2007xd}. They have also presented the probability density of the polarized quarks inside the pion and found that their spatial distribution is strongly distorted when the quarks are transversely polarized. The distortion in the density occurs due to the pion tensor FF. The lattice QCD results have triggered various theoretical studies on the pion and the kaon tensor FFs. The models for such results include constituent quark models~\cite{Frederico:2009fk,Fanelli:2016aqc}, the Nambu--Jona-Lasinio (NJL) model with Pauli-Villars regularization~\cite{Broniowski:2010nt,Dorokhov:2011ew}, and the nonlocal chiral quark model (N$\chi$QM) from the instanton vacuum~\cite{Nam:2010pt,Nam:2011yw}.

In this paper, we evaluate the GPDs of the light pseudoscalar mesons using the light-front wave functions (LFWFs) based on the theoretical framework of basis light front quantization (BLFQ) \cite{Vary:2009gt}, with only the valence Fock sector of mesons considered. The effective Hamiltonian incorporates the confining potential adopted from the light-front holography in the transverse direction \cite{Brodsky:2014yha}, a longitudinal confinement \cite{Li:2015zda,Li:2017mlw}, and the color-singlet NJL interactions~\cite{Klimt:1989pm,Shigetani:1993dx} to account for the dynamical chiral symmetry breaking of QCD. The nonperturbative solutions for the LFWFs are given by the recent BLFQ study  of light mesons~\cite{Jia:2018ary}. These  LFWFs have been applied successfully to  predict the decay constants, electromagnetic form factors (EMFFs), charge radii,  PDFs, and many other quantities of the pion and the kaon~\cite{Jia:2018ary, Lan:2019rba,Lan:2019vui,Mondal:2021czk}. Here, we extend those  investigations to study the pion and the kaon GPDs and their QCD evolution. 
We use the Dokshitzer-Gribov-Lipatov-Altarelli-Parisi (DGLAP) equation of QCD \cite{Dokshitzer:1977sg,Gribov:1972ri,Altarelli:1977zs} up to the next-to-next-to-leading order (NNLO) for the evolution of the valence quark GPDs.
We also calculate the pion and the kaon tensor FFs in the space-like  region. Combining the result of the tensor FFs with the EMFFs, which have been evaluated previously in Ref.~\cite{Jia:2018ary} within the BLFQ-NJL framework, we then compute the probability density of transversely polarized quarks inside the pion and the kaon. We further calculate the $x$-dependent squared radius of the quark density in the transverse plane that describes the transverse size of the hadron.

We organize the main results of this paper in the following sequence.  We briefly summarize the BLFQ-NJL
formalism for the light mesons in Sec.~\ref{sc:BLFQ_NJL}. We then present a  detailed description of the GPDs and the associated distributions in Sec.~\ref{formalism}. Sec.~\ref{result} details our numerical results for the GPDs, electromagnetic and gravitational FFs, impact parameter dependent GPDs,  and spin densities of the pion and the kaons. We summarize the outcomes in Sec.~\ref{summary}.

\section{BLFQ-NJL model for the light mesons}\label{sc:BLFQ_NJL}
In this section, we provide an overview of the BLFQ-NJL model for the light mesons following Ref.~\cite{Jia:2018ary}. The BLFQ approach represents the dynamics of bound state constituents in quantum field theory through a light-front quantum many-body Hamiltonian~\cite{Vary:2009gt,Zhao:2014xaa,Wiecki:2014ola,Li:2015zda,Jia:2018ary,Tang:2018myz,Tang:2019gvn,Xu:2019xhk,Lan:2021wok}. The structures of the bound states are encoded in the LFWFs achievable as the eigenfunctions of the light-front eigenvalue equation
\begin{equation}
H_{\mathrm{eff}}\vert \Psi\rangle=M^2\vert \Psi\rangle,\label{eq:LF_Schrodinger}
\end{equation}
where $H_{\mathrm{eff}}=P^+P^-$ with $P^\pm=P^0 \pm P^3$ being the light-front Hamitonian ($P^-$) and the longitudinal momentum ($P^+$) of the system, respectively.  The mass squared, $M^2$, is the corresponding eigenvalue of the state $\vert \Psi\rangle$.
In the constituent quark-antiquark representation, our adopted  effective light-front Hamiltonian for the light mesons with non-singlet flavor wave functions is written as
\begin{align}
H_\mathrm{eff} =& \frac{\vec k^{\perp2} + m_q^2}{x} + \frac{\vec k^{\perp2}+m_{\bar q}^2}{1-x}
+ \kappa^4 \vec \zeta^{\perp2}  \nonumber\\&- \frac{\kappa^4}{(m_q+m_{\bar q})^2} \partial_x\big( x(1-x) \partial_x \big)+H^{\rm eff}_{\rm NJL}.\label{eqn:Heff}
\end{align}
The first two terms in Eq.~\eqref{eqn:Heff} are the light-front kinetic energy for the quark and the antiquark,
where $m_q$ ($m_{\bar q}$) is the mass of the quark (antiquark), 
$x=k^+/P^+$ is the longitudinal momentum fraction carried by the valence quark, and 
$\vec{k}^{\perp}$ is its transverse momentum.  The third and the fourth terms are respectively the confining potential in the transverse direction based on the light-front holographic QCD~\cite{Brodsky:2014yha} and a longitudinal confining potential~\cite{Li:2015zda}. The parameter $\kappa$ is the strength of the confinement. The holographic variable is defined as~$\vec \zeta^{\perp} \equiv \sqrt{x(1-x)} \vec r^\perp$~\cite{Brodsky:2014yha}, where $\vec{r}^\perp$ is the transverse separation between the quark and antiquark and is conjugated to $\vec{k}^\perp$. The $x$-derivative is defined as $\partial_x f(x, \vec\zeta^\perp) = \partial f(x, \vec \zeta^\perp)/\partial x|_{\vec\zeta^\perp}$. The last term in the effective Hamiltonian, $H_{\mathrm{NJL}}^{\mathrm{eff}}$, represents the color-singlet NJL interaction to account for the chiral dynamics~\cite{Klimt:1989pm}.

For the positively-charged pion, the NJL interaction is given by~\cite{Jia:2018ary},
\begin{align}
&H_{\mathrm{NJL},\pi}^{\mathrm{eff}}  =G_\pi\, \big\{\overline{u}_{\mathrm{u}s1'}(p_1')u_{\mathrm{u}s1}(p_1)\,\overline{v}_{\mathrm{d}s2}(p_2)v_{\mathrm{d}s2'}(p_2')\nonumber\\
&\quad\quad+ \overline{u}_{\mathrm{u}s1'}(p_1')\gamma_5 u_{\mathrm{u}s1}(p_1)\,\overline{v}_{\mathrm{d}s2}(p_2)\gamma_5 v_{\mathrm{d}s2'}(p_2') \nonumber\\
&\quad\quad+ 2\,\overline{u}_{\mathrm{u}s1'}(p_1')\gamma_5 v_{\mathrm{d}s2'}(p_2')\,\overline{v}_{\mathrm{d}s2}(p_2)\gamma_5 u_{\mathrm{u}s1}(p_1) \big\}.\label{eq:H_eff_NJL_pi_ori}
\end{align}
While, for the positively charged kaon, the interaction is given by
\begin{align}
&H^{\mathrm{eff}}_{\mathrm{NJL},K}=G_K\,\big\{- 2\,\overline{u}_{\mathrm{u}s1'}(p_1') v_{\mathrm{s}s2'}(p_2')\,\overline{v}_{\mathrm{s}s2}(p_2) u_{\mathrm{u}s1}(p_1) \nonumber\\
&\quad\quad + 2\,\overline{u}_{\mathrm{u}s1'}(p_1')\gamma_5 v_{\mathrm{s}s2'}(p_2')\,\overline{v}_{\mathrm{s}s2}(p_2)\gamma_5 u_{\mathrm{u}s1}(p_1) \big\}.\label{eq:H_eff_NJL_SU_3_ori}
\end{align}
Equations~(\ref{eq:H_eff_NJL_pi_ori}) and (\ref{eq:H_eff_NJL_SU_3_ori}) are obtained from the NJL Lagrangian after the Legendre transform in the two and three flavor NJL model, respectively~\cite{Klimt:1989pm,Vogl:1989ea,Vogl:1991qt,Klevansky:1992qe}. Here, ${u_{\mathrm{f}s}(p)}$ and ${v_{\mathrm{f}s}(p)}$ are the Dirac spinors with the nonitalic subscripts representing the flavors and the italic subscripts denoting the spins. Meanwhile, $p_1$ and $p_2$ are the momenta of the valence quark and the valence antiquark, respectively.
The coefficients $G_{\pi}$ and $G_{K}$ are independent coupling constants of the theory. In the interactions, we only include the combinations of Dirac bilinears relevant to the valence Fock sector LFWFs of the systems. The instantaneous terms due to the NJL interactions have been omitted. The explicit expressions and the detailed calculations of the matrix elements of the NJL interactions in the BLFQ formalism can be found in Ref.~\cite{Jia:2018ary}. 

In the leading Fock sector, the eigenstate for the mesons reads  
\begin{align}
 &\big\vert\Psi(P^+,\vec{P}^\perp)\big\rangle =\sum_{r,s}\int_{0}^{1}\dfrac{dx}{4\pi x(1-x)}\int\dfrac{d\vec{\kappa}^\perp}{(2\pi)^2}\,\nonumber\\
	 &\quad\quad\times\,\psi_{rs}(x,\vec{\kappa}^\perp)\, b_r^\dagger(xP^+,\vec{\kappa}^\perp+x\vec{P}^\perp) \nonumber\\
	 &\quad\quad\times\,
	d_s^\dagger((1-x)P^+,-\vec{\kappa}^\perp+(1-x)\vec{P}^\perp)\,|0\rangle,\label{eq:Psi_meson_qqbar}
\end{align}
where $P$ is the momentum of the meson. The relative transverse momentum of the valence quark is $\vec{\kappa}^\perp=\vec{k}^\perp-x\vec{P}^\perp$. The coefficients of the expansion, $\psi_{rs}(x,\vec{\kappa}^\perp)$, are the valence sector LFWFs with $r$($s$) representing the spin of the quark(antiquark).
To compute the Hamiltonian matrix, one needs to construct the BLFQ basis. The two-dimensional (2D) harmonic oscillator (HO) basis functions are adopted in the transverse direction, which are defined as~\cite{Vary:2009gt,Li:2015zda}:
\begin{align}
	\phi_{nm}\left(\vec{q}^\perp;b_h \right)& =\dfrac{1}{b_h}\sqrt{\dfrac{4\pi n!}{(n+|m|)!}} \left(\dfrac{\vert\vec{q}^\perp\vert}{b_h}\right)^{|m|} \nonumber\\
	 &\times\, \exp\left(-\dfrac{\vec{q}^{\perp 2}}{2b_h^2}\right)
	  L_n^{|m|} \left(\dfrac{\vec{q}^{\perp 2}}{b_h^2}\right)\,e^{im\varphi},\label{eq:def_phi_nm}
	\end{align}
	with $\tan(\varphi)=q^2/q^1$, $b_h$ is the HO basis scale parameter with dimension of mass, $n$ and $m$ are the radial and the angular quantum numbers, $L_n^{|m|}(z)$ is the associated Laguerre polynomial. Meanwhile, in the longitudinal direction, the basis functions are defined as~\cite{Li:2015zda}
	\begin{align}
	\chi_l(x;\alpha,\beta)&= \sqrt{4\pi(2l+\alpha+\beta+1)}\nonumber\\
	 &\times\,\sqrt{\dfrac{\Gamma(l+1)\Gamma(l+\alpha+\beta+1)}{\Gamma(l+\alpha+1)\Gamma(l+\beta+1)}} \nonumber\\
	 &\times\, x^{\beta/2}(1-x)^{\alpha/2}\,P_l^{(\alpha,\beta)}(2x-1),\label{eq:def_chi_l}
	\end{align}
	where $P_{l}^{(\alpha,\beta)}(z)$ is the Jacobi polynomial and the dimensionless parameters ${\alpha =2m_{\overline{q}}(m_q+m_{\overline{q}})/\kappa^2}$, ${\beta=2m_q(m_q+m_{\overline{q}})/\kappa^2}$ and $l=0,~1,~2,...$. 
The valence LFWFs are  then expanded in the orthonormal bases given in Eqs.~(\ref{eq:def_phi_nm}) and (\ref{eq:def_chi_l}):
\begin{align} 
	 \psi_{rs}(x,\vec{\kappa}^\perp) &=\sum_{n, m, l}  \langle n, m, l, r, s | \psi\rangle~ \nonumber\\
	 &\times\,\phi_{nm}\left(\dfrac{\vec{\kappa}^\perp}{\sqrt{x(1-x)}};b_h\right)\chi_l(x),\label{eq:psi_rs_basis_expansions}
\end{align}
where the coefficients $\langle n, m, l,r,s|\psi\rangle$ are obtained in the BLFQ basis space by diagonalizing the truncated Hamiltonian matrix.
The infinite dimensional basis is truncated to a finite dimension by restricting the quantum numbers using
	\begin{equation}
	0 \leq n \leq N_{\mathrm{max}}, \quad -2 \leq m \leq 2, \quad 0 \leq l \leq L_{\mathrm{max}},
	\label{eq:nmax}
	\end{equation}
where $N_{\text{max}}$ controls the transverse momentum covered by 2D HO functions and $L_{\text{max}}$ provides the basis resolution in the longitudinal direction.
Note that we have a natural truncation for $m$ as the NJL interactions do not couple to $\vert m\vert \geq 3$ basis states~\cite{Jia:2018ary}.
%
The LFWF $\psi_{rs}(x,\vec{\kappa}^\perp)$ is normalized as
\begin{align}
\sum_{r,s}
\int_0^1 \!\! \frac{dx}{2x(1-x)} \!\int \! \frac{d^2 \vec{\kappa}^\perp}{(2\pi)^3}  \big | \psi_{rs}(x,\vec{\kappa}^\perp)\big |^2 \!\! =\!\!1. \label{eq:normalization_LFWF}
\end{align}

Parameters in the BLFQ-NJL model are fixed to reproduce the ground state masses of the light pseudoscalar and vector mesons as well as the experimental charge radii of the $\pi^+$ and the $K^+$~\cite{Jia:2018ary}. The LFWFs in this model have been successfully applied to compute the parton distribution amplitudes and the EMFFs \cite{Jia:2018ary}, PDFs for the pion and the kaon and pion-nucleus induced Drell-Yan cross sections~\cite{Lan:2019vui,Lan:2019rba}.

 \begin{table}
\caption{Summary of the model parameters~\cite{Jia:2018ary}.
}\label{tab:model_parameters}
 \centering
\begin{tabular}{ccc ccc ccc c}
\toprule
Valence flavor & $N_{\text{max}}$ &  $L_{\text{max}}$   &  $\kappa (\text {MeV})$  & $m_q (\text {MeV})$   & $m_{\bar {q}} (\text {MeV})$  &   \\

\colrule

$u\bar{d}$ & 8 & 8$-$32 & 227  & 337 & 337  &    \\

\colrule

$u\bar{s}$ & 8& 8$-$32 & 276   & 308 & 445  &    \\

\botrule
\end{tabular}
\end{table}
\begin{figure*}
\begin{tabular}{cc}
\subfloat[]{\includegraphics[scale=0.57]{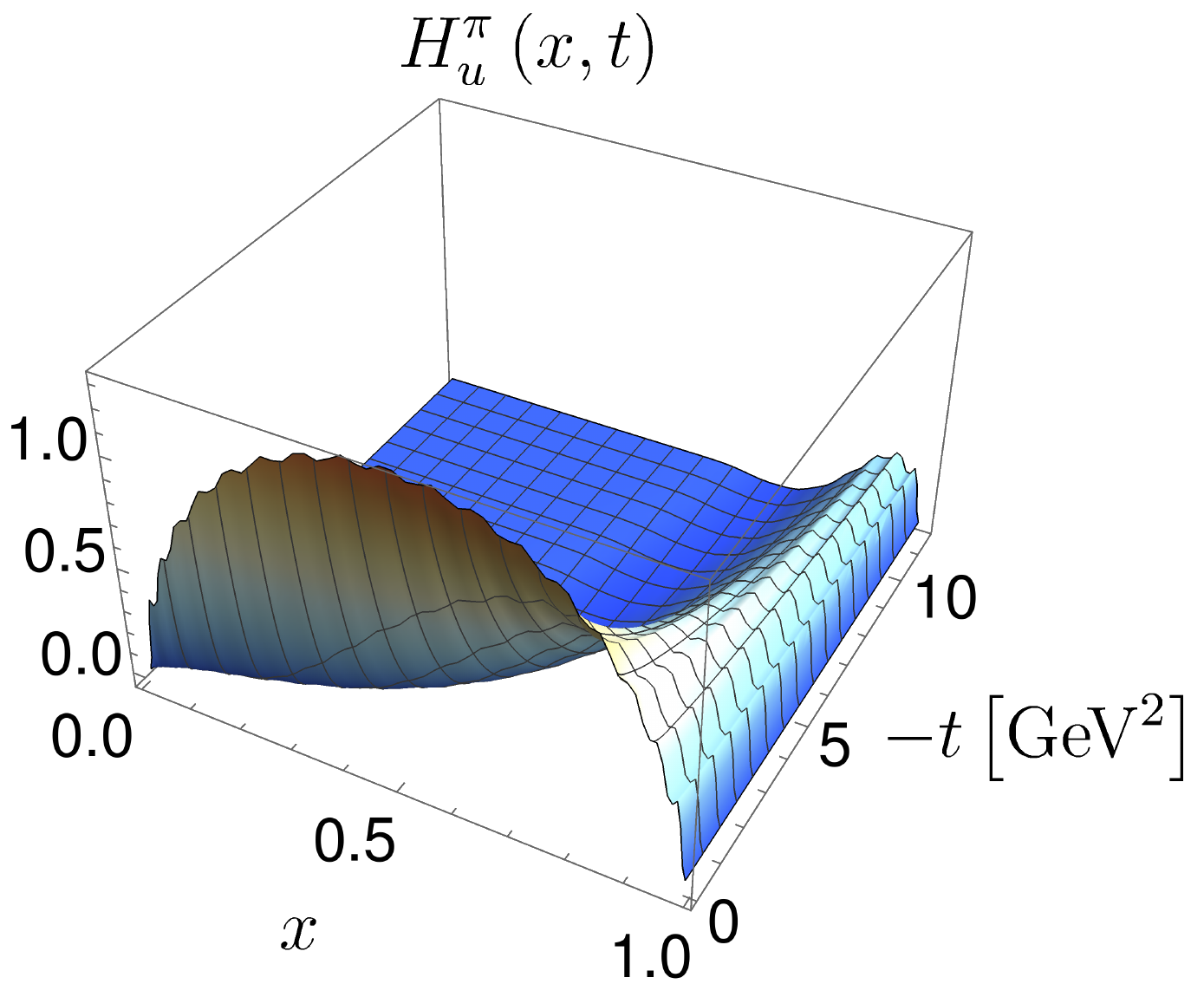}}
\end{tabular}
\begin{tabular}{cc}
\subfloat[]{\includegraphics[scale=0.57]{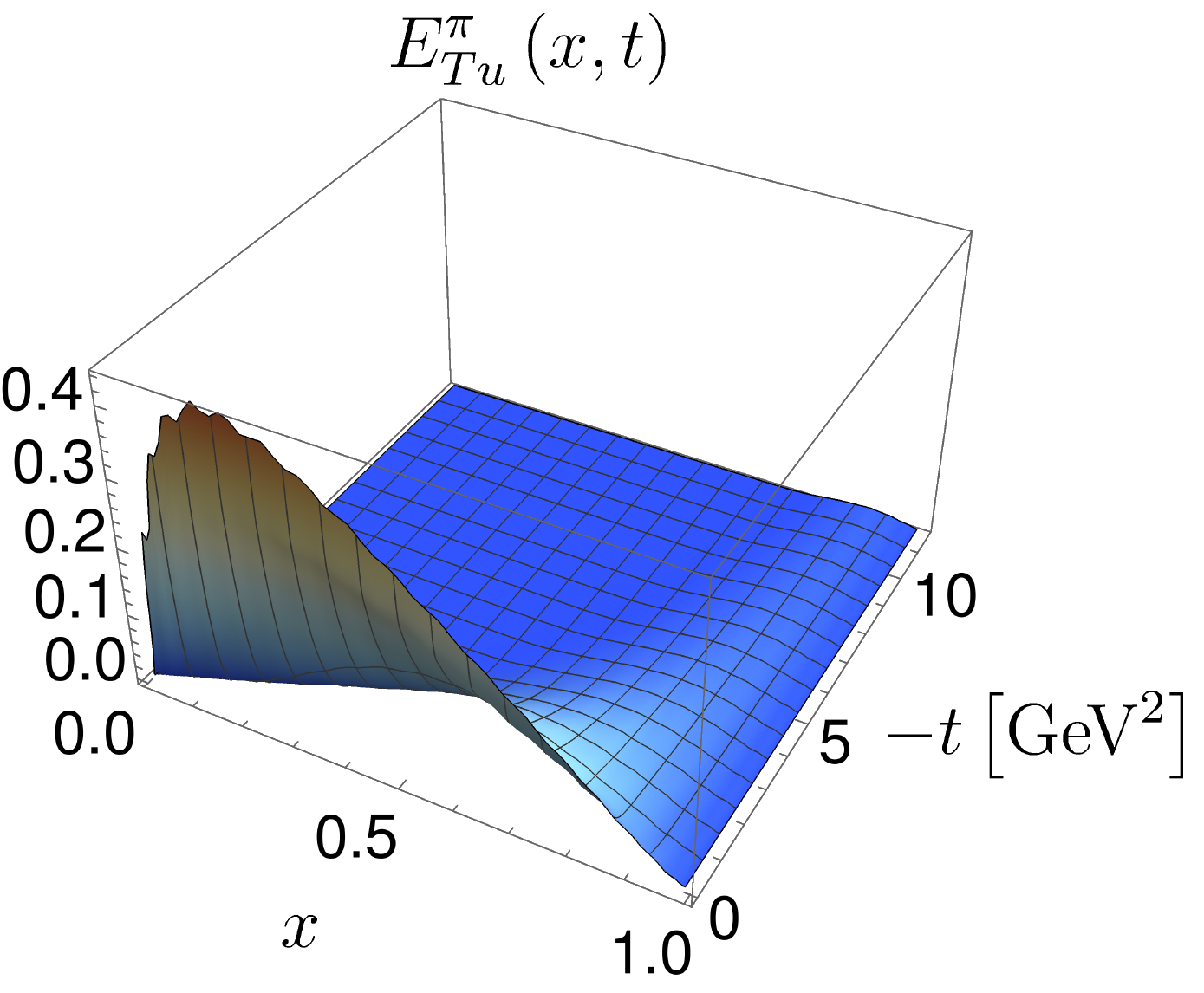}}
\end{tabular}
\caption{The valence ($u$ or $\bar{d}$) quark GPDs of the pion: (a) $H(x,0,t)$ and (b) $E_T(x,0,t)$ as functions of $x$ and the invariant momentum transfer $-t$. The GPDs are evaluated with $N_\text{max} = 8$ and $L_\text{max} = 32$ in the BLFQ-NJL model.}
\label{pion_gpds}
\end{figure*}
\begin{figure*}
\begin{tabular}{cc}
\subfloat[]{\includegraphics[scale=0.57]{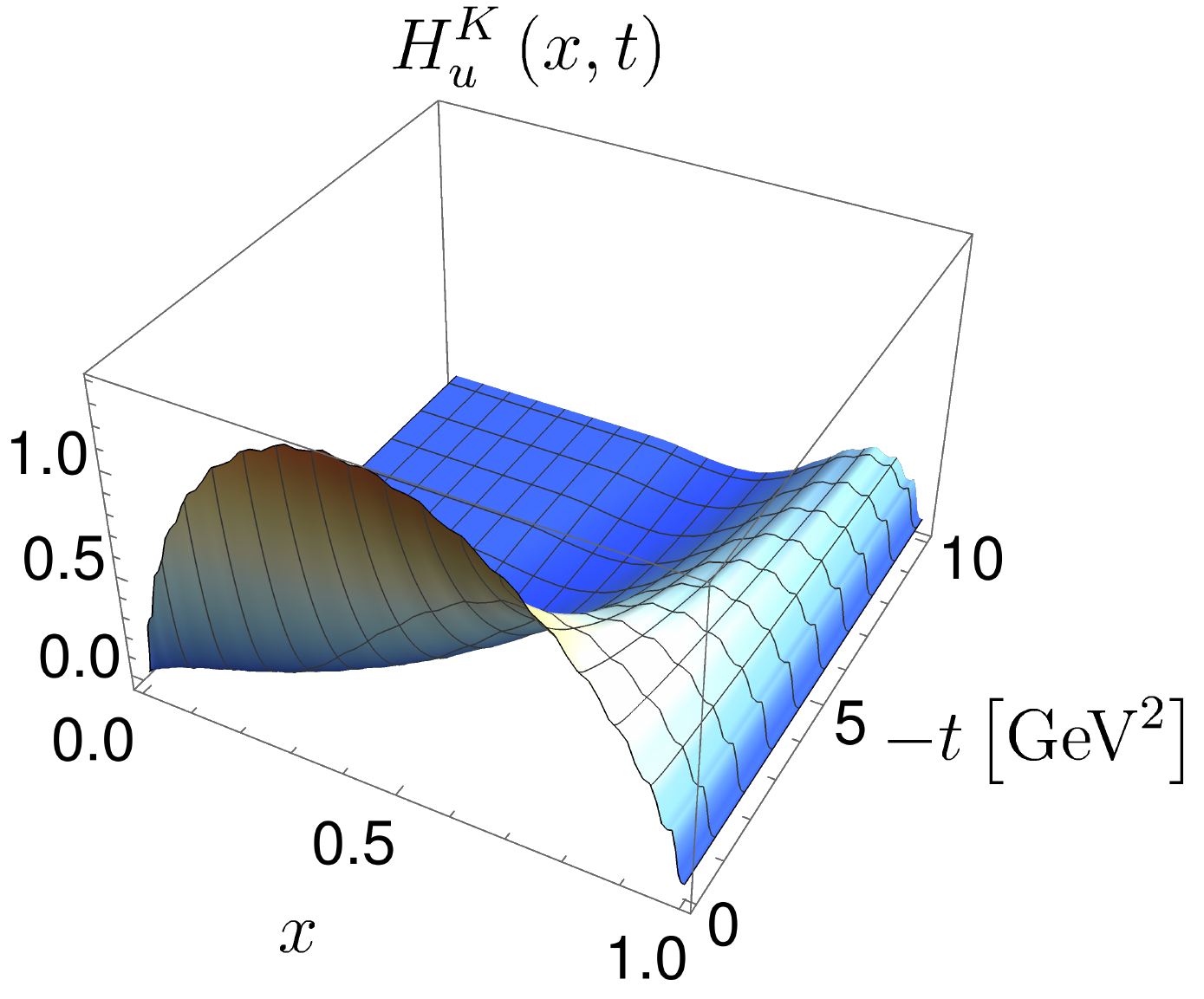}}
\end{tabular}
\begin{tabular}{cc}
\subfloat[]{\includegraphics[scale=0.57]{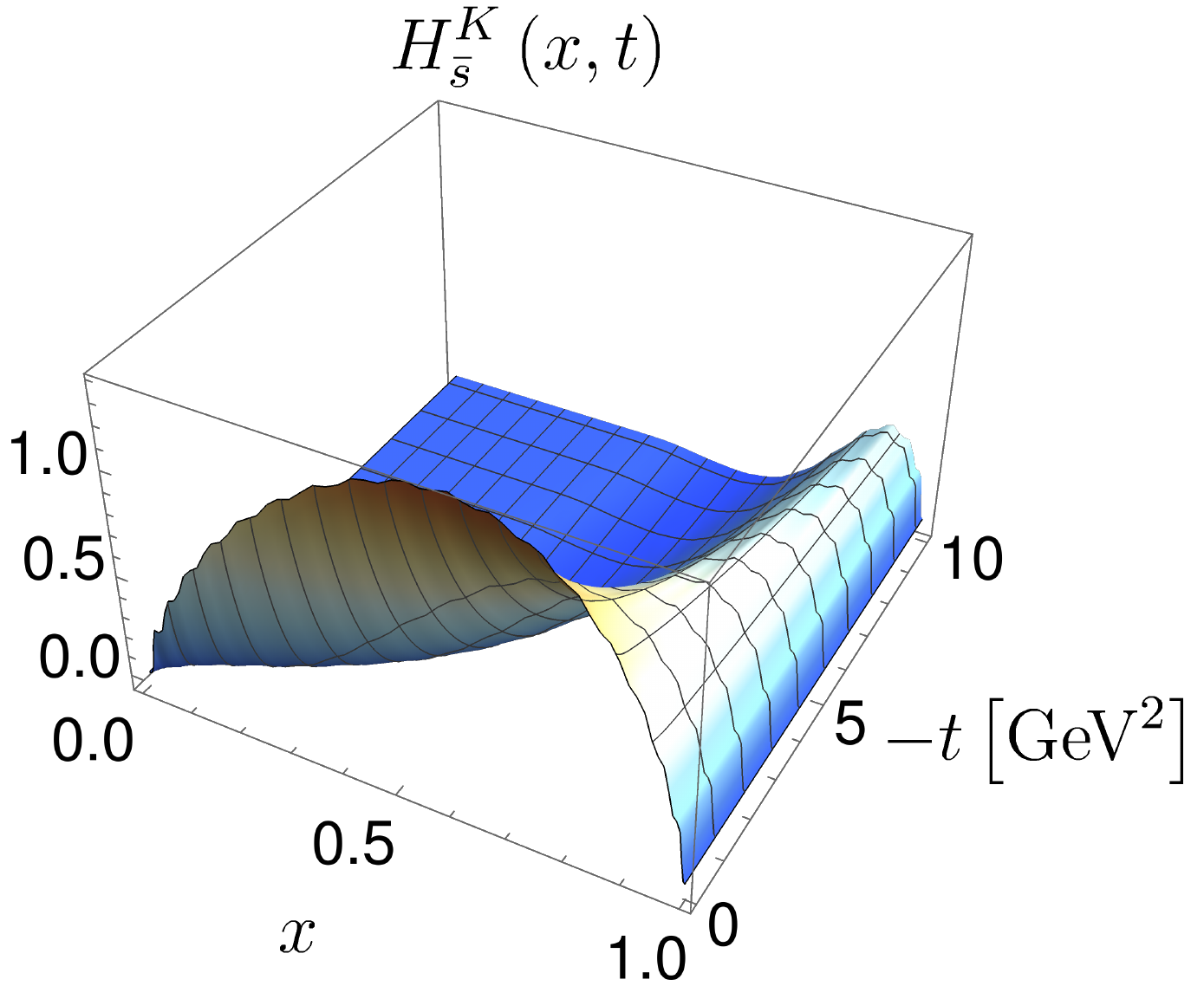}}
\end{tabular}
\begin{tabular}{cc}
\subfloat[]{\includegraphics[scale=0.57]{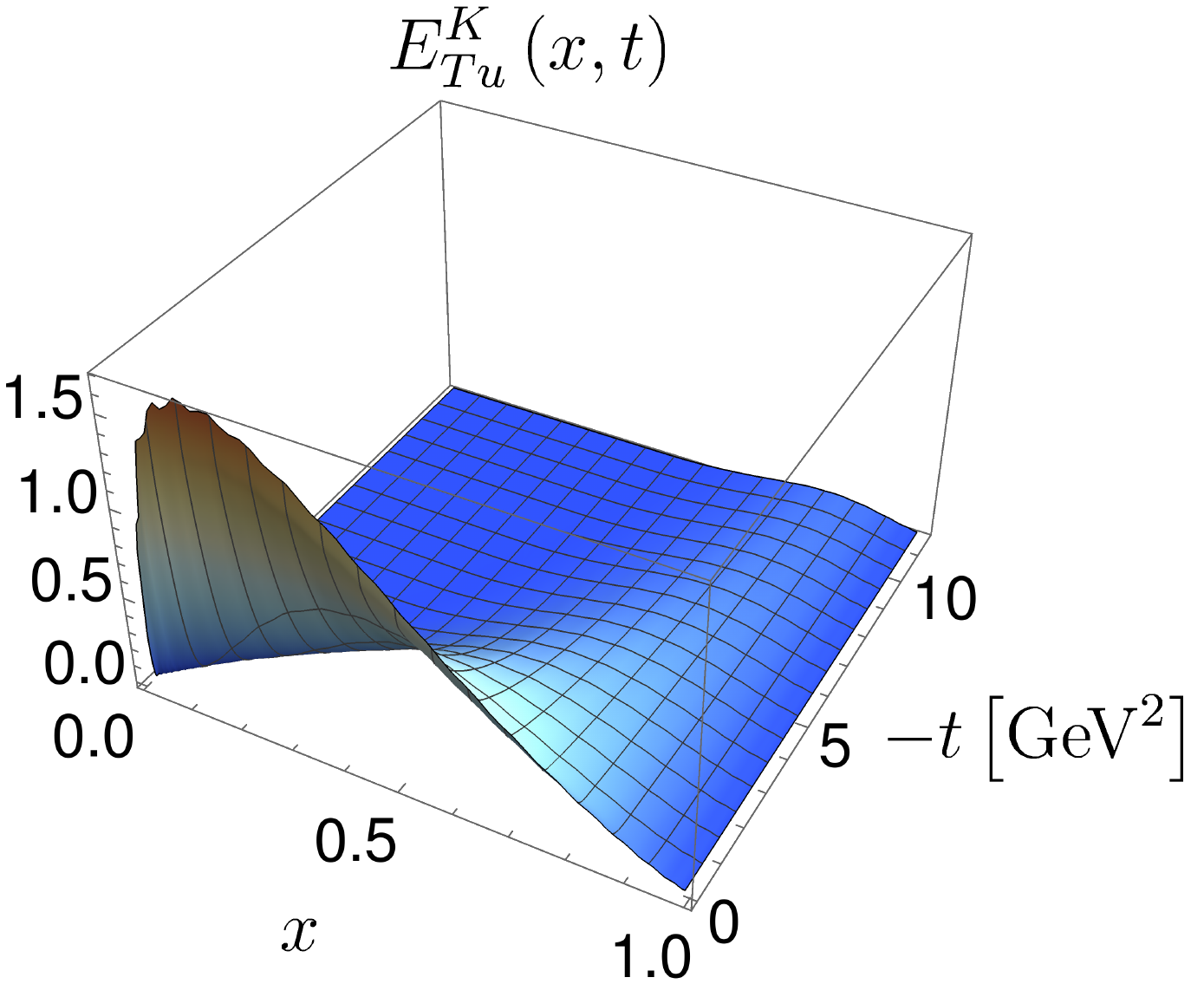}}
\end{tabular}
\begin{tabular}{cc}
\subfloat[]{\includegraphics[scale=0.57]{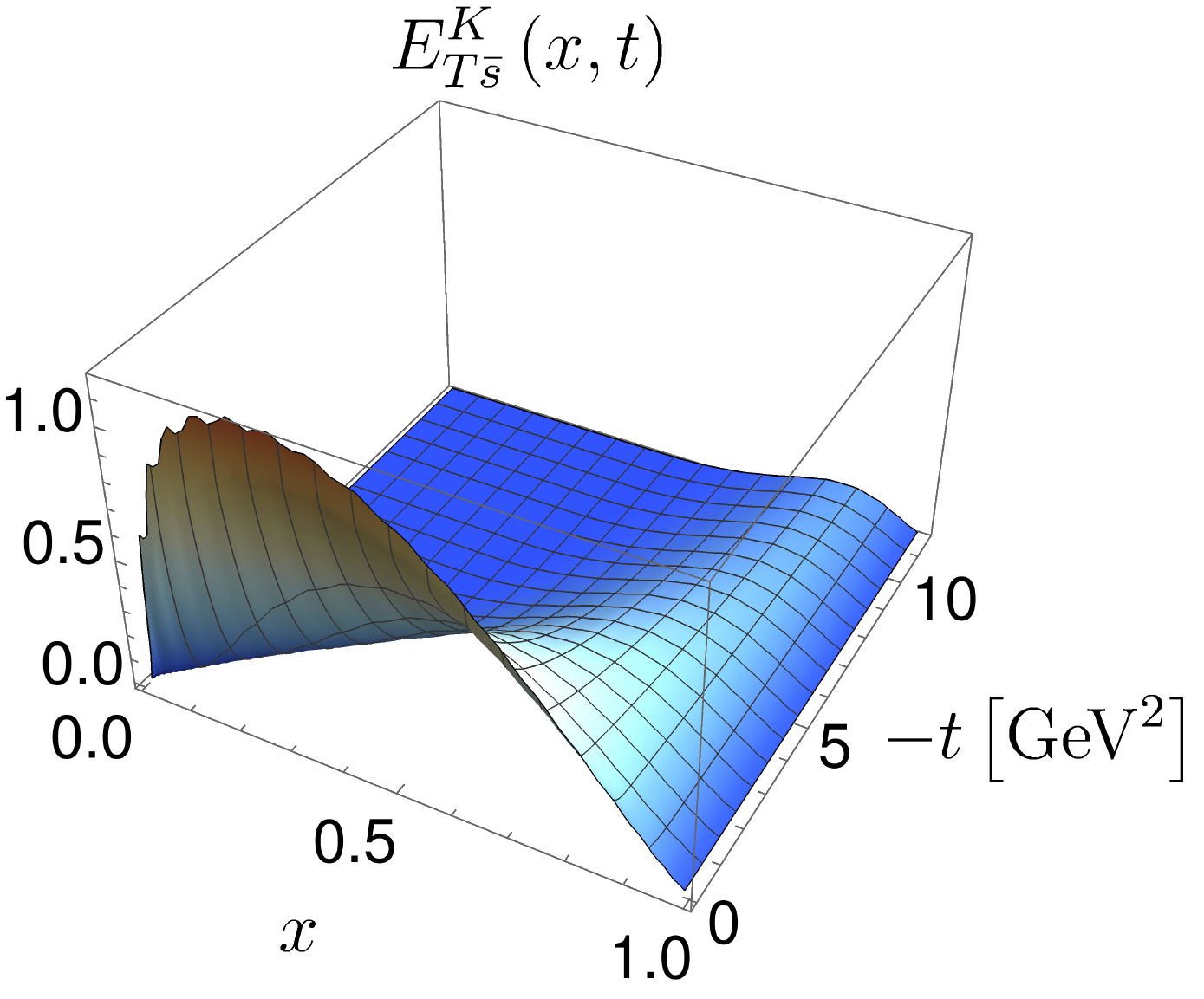}}
\end{tabular}
\caption{The valence quark GPDs of the kaon: (a) $H(x,0,t)$ and (c) $E_T(x,0,t)$ are for the valence $u$ quark; (b) and (d) are same as (a) and (c), respectively, but for the valence $\bar{s}$ quark as functions of $x$ and the invariant momentum transfer $-t$. These GPDs are evaluated  with $N_\text{max} = 8$ and $L_\text{max} = 32$ in the BLFQ-NJL model.}
\label{kaon_gpds}
\end{figure*}
\section{generalized parton distributions: KINEMATICS AND FORMALISM \label{formalism}}
At leading twist, there are two independent GPDs for a spin-0 meson. One of them is chirally even, while the other is chirally odd.
Those GPDs are defined as off-forward matrix elements of the bilocal operator of light-front correlation functions of vector and tensor currents, respectively as~\cite{Diehl:2003ny,Hagler:2009ni},
\begin{align}
&H^{\mathcal{P}}(x,\zeta,t)= \int \frac{dz^-}{4 \pi} e^{i x P^+ z^-/2}\nonumber\\
	 &\quad\times\,\langle\mathcal{P}(P')\vert\bar{\Psi}_q(0)\gamma^+ \Psi_q(z)\vert\mathcal{P}(P)\rangle\vert_{z^+=\textbf{z}^\perp=0},\label{H}\\
&\frac{i \epsilon_{ij}^\perp q_i^\perp}{2M_\mathcal{P}}E_T^\mathcal{P}(x,\zeta,t)= \int \frac{dz^-}{4 \pi} e^{i x P^+ z^-/2}\nonumber\\
	 &\quad\times\,\langle\mathcal{P}(P')\vert\bar{\Psi}_q(0)i \sigma^{j+}\gamma_5 \Psi(z)_q\vert\mathcal{P}(P)\rangle\vert_{z^+=\textbf{z}^\perp=0},\label{ET}
\end{align} 
 where $\Psi_q(z)$ is the quark field operator and $P(P')$ denotes the meson momentum of initial (final) state of the meson $(\mathcal{P})$. $M_{\mathcal{P}}$ defines the mass of the meson; $\epsilon_{ij}^\perp$ is the anti-symmetric tensor in the transverse plane and $\sigma^{j+}=\frac{i}{2}[\gamma^j,\gamma^+]$ with $j=1,~2$ as transverse index. The $H$, called the  unpolarized quark GPD, is chirally even, while the transversely polarized quark GPD, $E_T$, is chirally odd. The GPD $E_T$ is responsible for the distortion in the spatial distribution of a transversely polarized quark, revealing a nontrivial spin structure of the meson~\cite{Brommel:2007xd}. The moments of the GPD $E_T$ can be linked to the Boer-Mulders function, which describes the correlation between transverse spin and intrinsic transverse momentum of the quark in the meson~\cite{Burkardt:2003uw,Burkardt:2002ks,Ahmady:2019yvo}. Recently, the limits of validity of this relationship have been discussed in Ref.~\cite{Pasquini:2019evu}. In the symmetric frame, the kinematical variables are 
\begin{equation}
\bar{P}^\mu=\frac{(P+P')^\mu}{2}, ~ \Delta^\mu=P'^\mu-P^\mu, ~ \zeta=-\Delta^+/2\bar{P}^+,
\end{equation}
and $t=\Delta^2$. Here, we choose the light cone gauge $A^+=0$ implying that the gauge link between the quark fields in Eqs.~(\ref{H}) and (\ref{ET}) is unity therefore omitted.

By inserting the initial and the final states of the meson, Eq.~(\ref{eq:Psi_meson_qqbar}), in above Eqs.~(\ref{H}) and (\ref{ET}), one obtains the quark GPDs $H$ and $E_T$ in terms of overlaps of LFWFs. We restrict ourselves  to the kinematical region: $0<x<1$ at zero skewness. This domain corresponds to the situation where a quark is removed from the initial meson with light-front
longitudinal momentum $xP^+$ and reinserted into the final meson with the same longitudinal momentum. Therefore, the
change in momentum occurs purely in the transverse
direction. The particle number $(n_p)$ remains conserved in this kinematical domain describing the diagonal $n_p\to n_p$ overlaps. The GPDs, $H$ and $E_T$, at zero skewness, in the diagonal $2 \to 2$ overlap representation in terms of LFWFs are given by 
\begin{align}
 H(x,\zeta=0,t)&=\dfrac{1}{4\pi\,x(1-x)}\sum_{rs}\int \dfrac{d^2\vec{\kappa}^\perp}{(2\pi)^2}\,\nonumber\\
	 &\times\,\psi^*_{rs}(x,\vec{\kappa}'^\perp)\,\psi_{rs}(x,\vec{\kappa}^\perp),\label{eq:H_valence_psi}\\
 \frac{i \Delta^\perp_j}{2M_{\mathcal{P}}}E_T(x,\zeta=0,t)&=\dfrac{1}{4\pi\,x(1-x)}\sum_{s}\int \dfrac{d^2\vec{\kappa}^\perp}{(2\pi)^2}\,\nonumber\\
	 &\times\,\Big[(-i)^j\psi^*_{\uparrow s}(x,\vec{\kappa}'^\perp)\,\psi_{\downarrow s}(x,\vec{\kappa}^\perp)\nonumber\\
 &  +(i)^j\psi^*_{\downarrow s}(x,\vec{\kappa}'^\perp)\,\psi_{\uparrow s}(x,\vec{\kappa}^\perp)\Big],\label{eq:ET_valence_psi}
\end{align}
 where, for the struck quark, $\vec{\kappa}'^\perp=\vec{\kappa}^\perp+(1-x){\vec \Delta}^\perp$ and for the spectator, $\vec{\kappa}'^\perp=\vec{\kappa}^\perp-x{\vec \Delta}^\perp$ and  the total momentum transferred to the meson is $t=-{\vec \Delta}^{\perp 2}$. 

Note that integrating the bilocal matrix element in Eq.~(\ref{H}) over the momentum fraction $x$ yields the local matrix elements that provide FFs. In the Drell-Yan frame, the expressions for the GPDs are very similar to those for FFs, except that the longitudinal momentum fraction $x$ of the struck parton is not integrated out. Therefore, GPDs defined in Eqs.~(\ref{eq:H_valence_psi}) and (\ref{eq:ET_valence_psi}) are also known as momentum-dissected FFs and measure the contribution of the struck parton with momentum fraction $x$ to the corresponding FFs. Consequently, the first moments of the GPDs can be related to the FFs for the spin-0 hadrons by the sum rules on the light-front as~\cite{Ji:1998pc}
\begin{align}
 F(t)&=\int dx \, H(x,\zeta,t),\nonumber\\	
 F_T(t)&=\int dx \, E_T(x,\zeta,t).\label{eq:FF}
\end{align}
Meanwhile, the gravitational FFs which are expressed as
the matrix elements of the energy-momentum tensor, are linked to GPDs through the second-moment as~\cite{Ji:1998pc}
\begin{align}
 A(t)&=\int dx \, x\,H(x,\zeta,t),\nonumber\\
 B_T(t)&=\int dx \, x\, E_T(x,\zeta,t).\label{eq:GF}
\end{align}

Aside from these FFs, the impact parameter dependent GPDs are defined as the Fourier transform of the GPDs with respect to the momentum transfer along the transverse direction $\vec{\Delta}^\perp$~\cite{Burkardt:2002hr}:
\begin{align}
  q(x, {\vec b}^\perp)& =
\int \frac{d^2{\vec \Delta}^\perp}{(2\pi)^2}
e^{-i {\vec \Delta}^\perp \cdot {\vec b}^\perp }
H(x,0,-{\vec \Delta}^{\perp 2}),\label{eq:Hb}\\
q_T(x, {\vec b}^\perp)& =\int \frac{d^2{\vec \Delta}^\perp}{(2\pi)^2}
e^{-i {\vec \Delta}^\perp \cdot {\vec b}^\perp }
E_T(x,0,-{\vec \Delta}^{\perp 2}),\label{eq:Eb}
\end{align}
where ${\vec b}^\perp$ is the Fourier conjugate to the momentum transfer ${\vec \Delta}^\perp$. The impact parameter ${b}^\perp=|{\vec b}^\perp|$ corresponds to the transverse displacement of the struck parton from the center of momentum  of the hadron. For zero skewness, $\vec{b}^{\perp}$ provides a measure of the transverse distance of the struck parton from the center of momentum of the hadron. The variable $\vec{b}^{\perp}$ follows the condition $\sum_i x_i \vec{b}^{\perp}_{i}=0$, where the sum runs over the number of partons. The relative distance between the center of momentum of the spectator and the struck parton is ${{b}^{\perp}/(1-x})$, therefore providing an estimate of the transverse size of the hadron \cite{Diehl:2003ny}.

Following the standard formulation \cite{Miller:2007uy},  one can further define the transverse charge density $\rho({\vec b}^\perp)$ by
\begin{align}
  \rho({\vec b}^\perp)&= \int \frac{d^2{\vec \Delta}^\perp}{(2\pi)^2}
e^{-i {\vec \Delta}^\perp \cdot {\vec b}^\perp } F(-{\vec \Delta}^{\perp 2})\nonumber\\&=\int_0^1 dx \,  q(x, {\vec b}^\perp)\label{eq:rho},
\end{align}
while the longitudinal momentum density for a given transverse seperation is given by~\cite{Abidin:2008sb,Chakrabarti:2015lba,Mondal:2015fok,Kumar:2017dbf,Mondal:2016xsm}
\begin{align}
  p({\vec b}^\perp)&= \int \frac{d^2{\vec \Delta}^\perp}{(2\pi)^2}
e^{-i {\vec \Delta}^\perp \cdot {\vec b}^\perp } A(-{\vec \Delta}^{\perp 2})\nonumber\\&=\int_0^1 dx \, x\, q(x, {\vec b}^\perp)\label{eq:rho2}.
\end{align}
\begin{figure*}
\begin{tabular}{cc}
\subfloat[]{\includegraphics[scale=0.3]{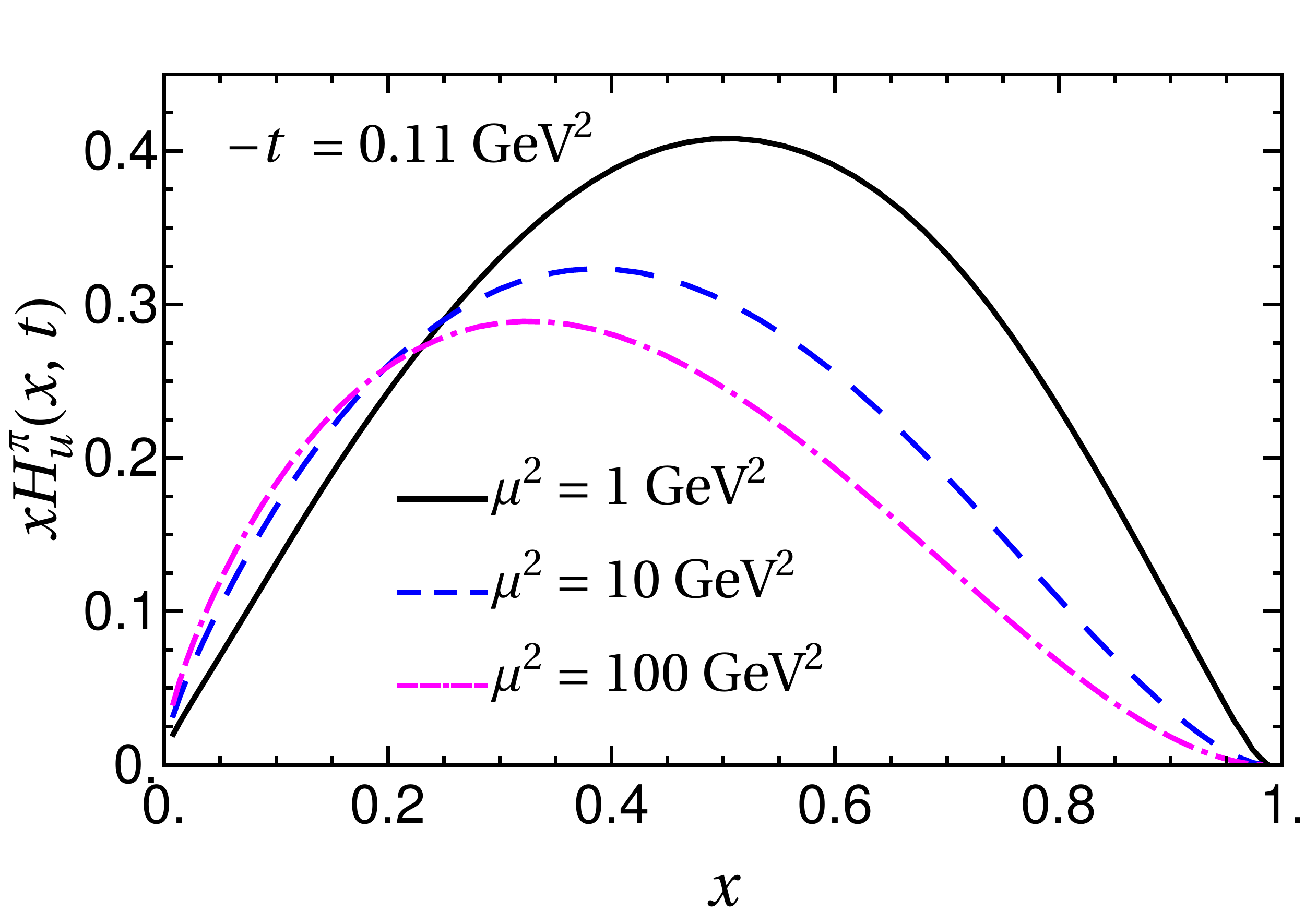}}
\end{tabular}
\begin{tabular}{cc}
\subfloat[]{\includegraphics[scale=0.3]{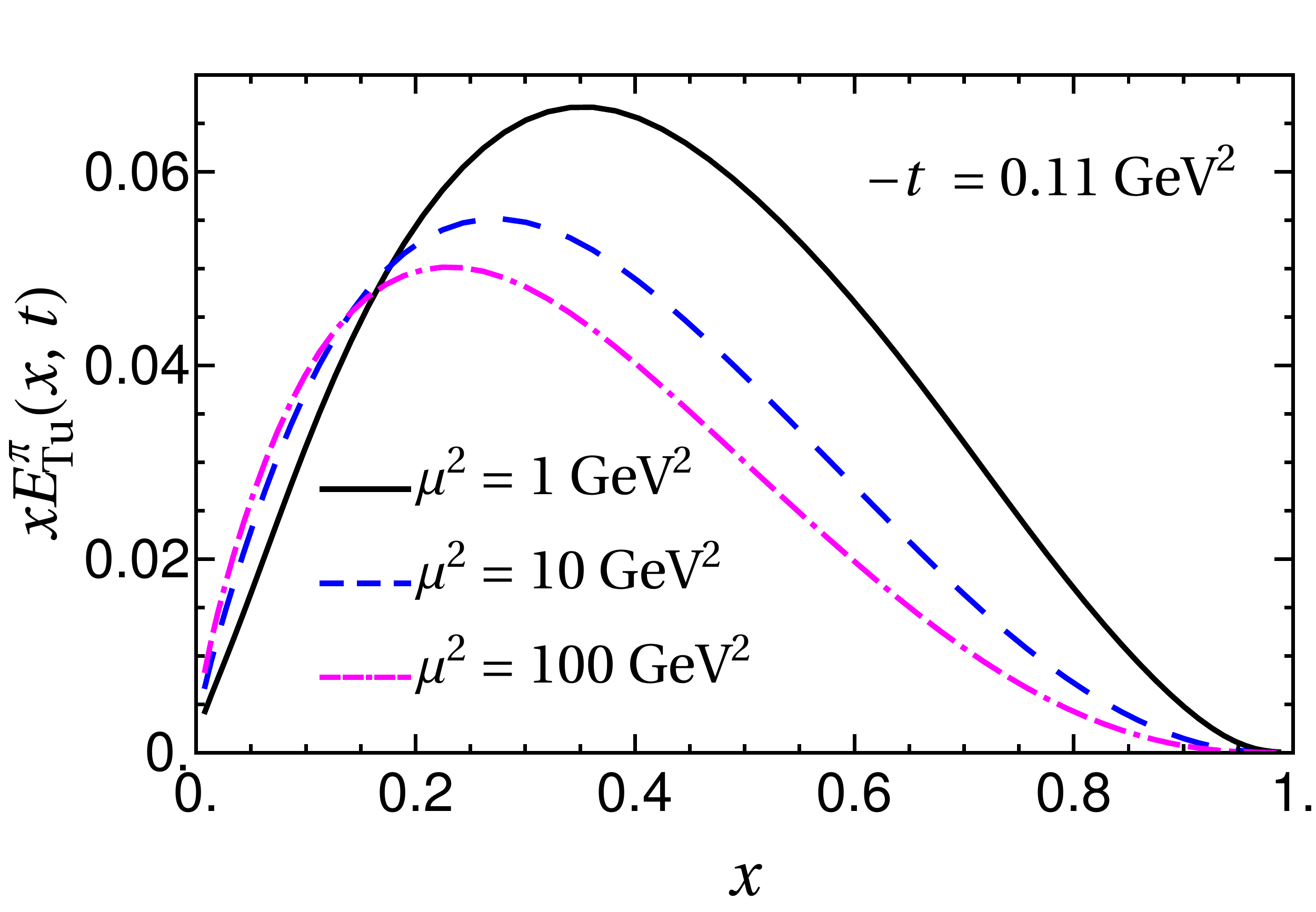}}
\end{tabular}
\caption{Scale evolution of the valence ($u$ or $\bar{d}$) quark GPDs of the pion: (a) $H(x,0,t)$ and (b) $E_T(x,0,t)$ as functions of $x$ and fixed value of $-t=0.11$ GeV$^2$. The GPDs are evaluated with $N_\text{max} = 8$ and $L_\text{max} = 32$  within the BLFQ-NJL model. The GPDs are evolved from our model scale for the pion $\mu_{0\pi}^2=0.240$ GeV$^2$ to the final scales $\mu^2=1,~10,~100$ GeV$^2$.}
\label{evolution_pion_gpds}
\end{figure*}
\begin{figure*}
\begin{tabular}{cc}
\subfloat[]{\includegraphics[scale=0.3]{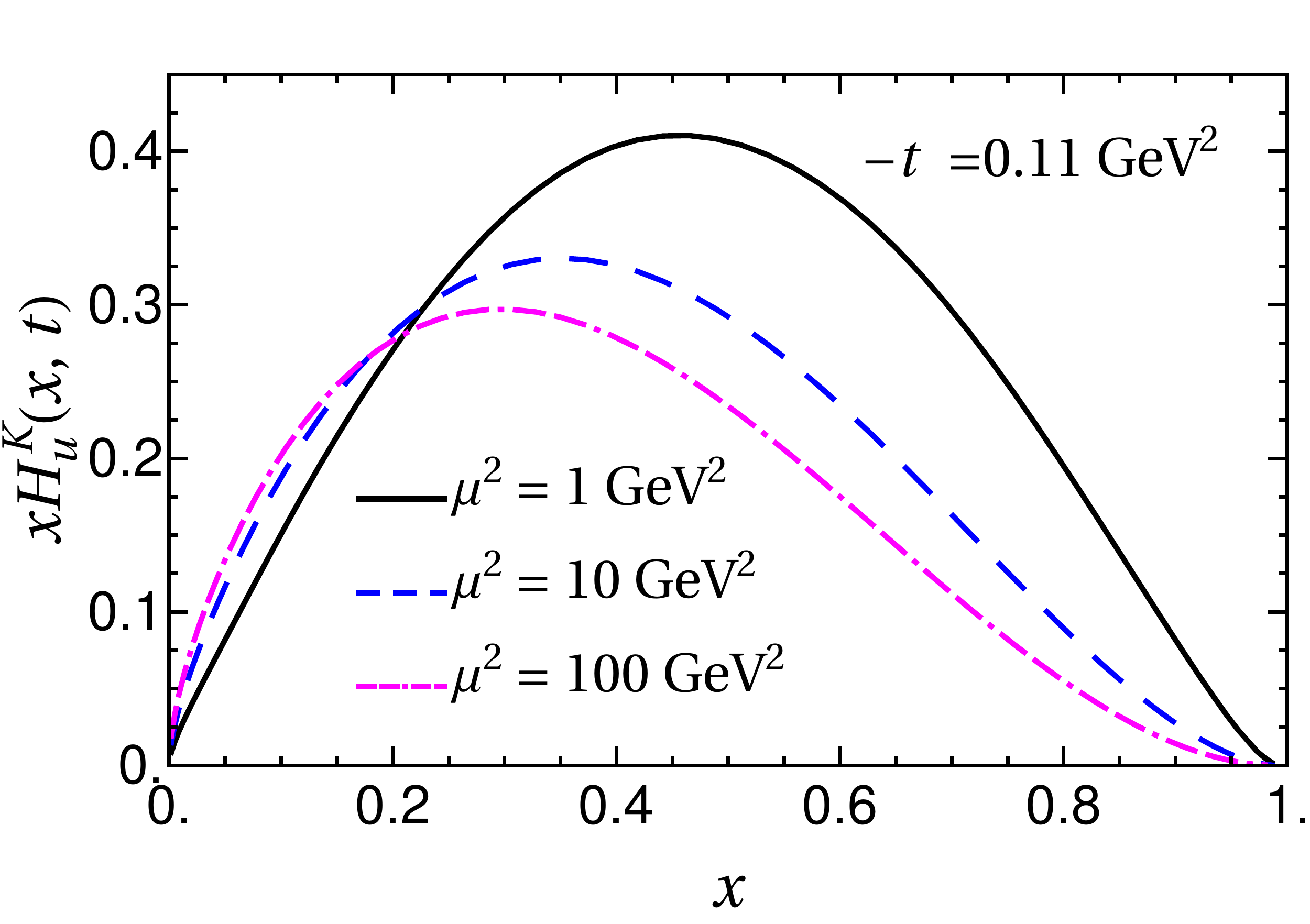}}
\end{tabular}
\begin{tabular}{cc}
\subfloat[]{\includegraphics[scale=0.3]{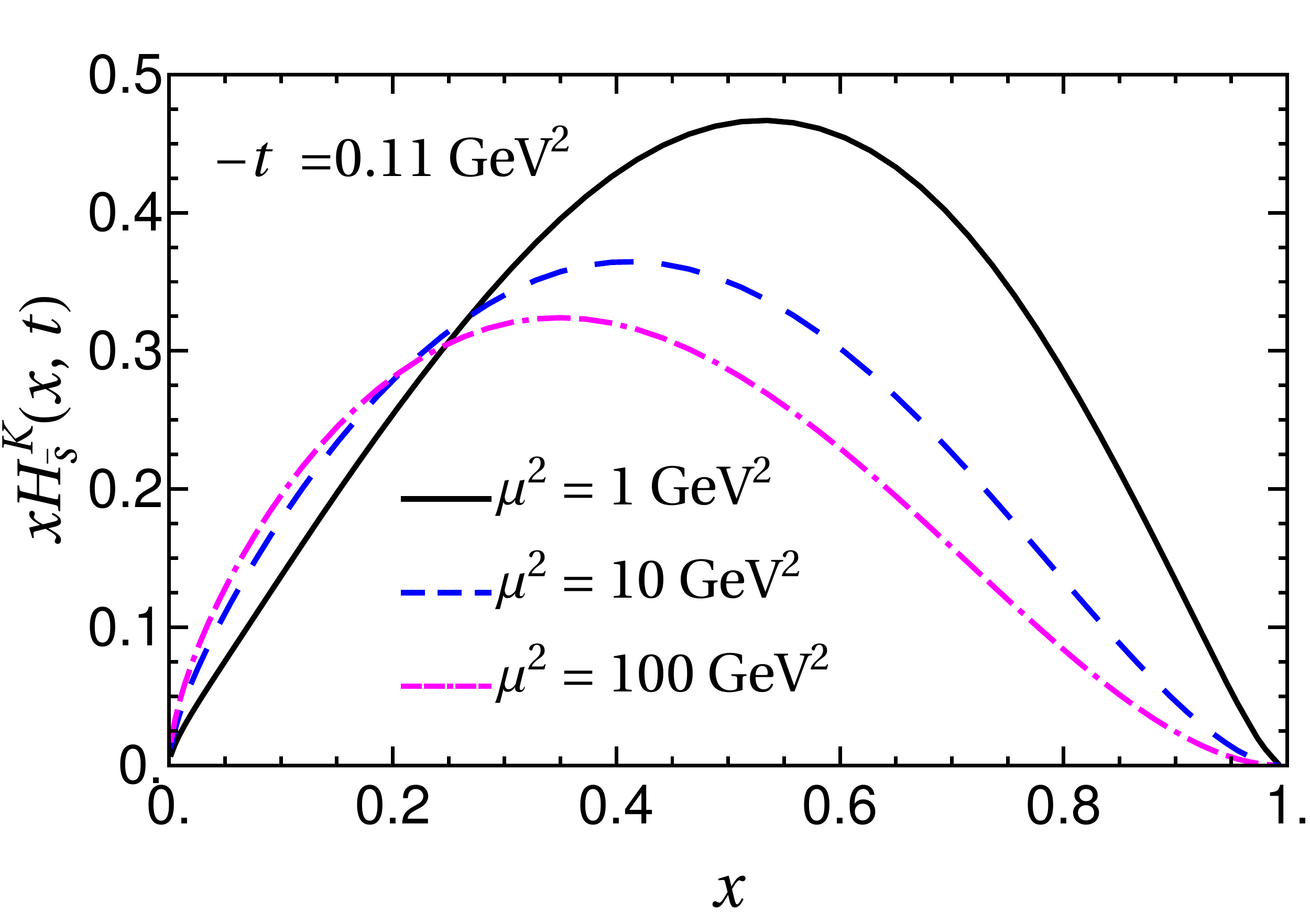}}
\end{tabular}
\begin{tabular}{cc}
\subfloat[]{\includegraphics[scale=0.3]{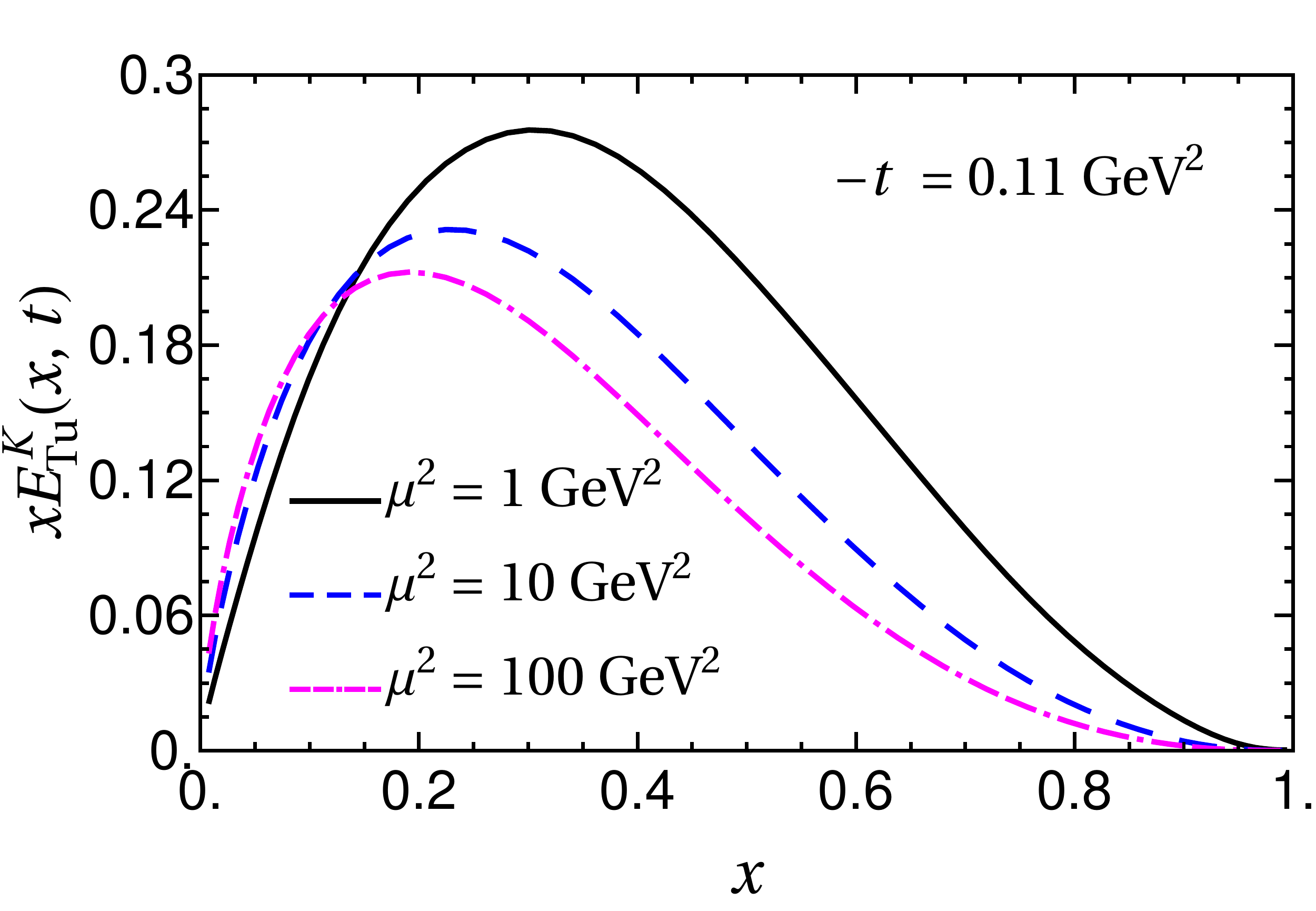}}
\end{tabular}
\begin{tabular}{cc}
\subfloat[]{\includegraphics[scale=0.3]{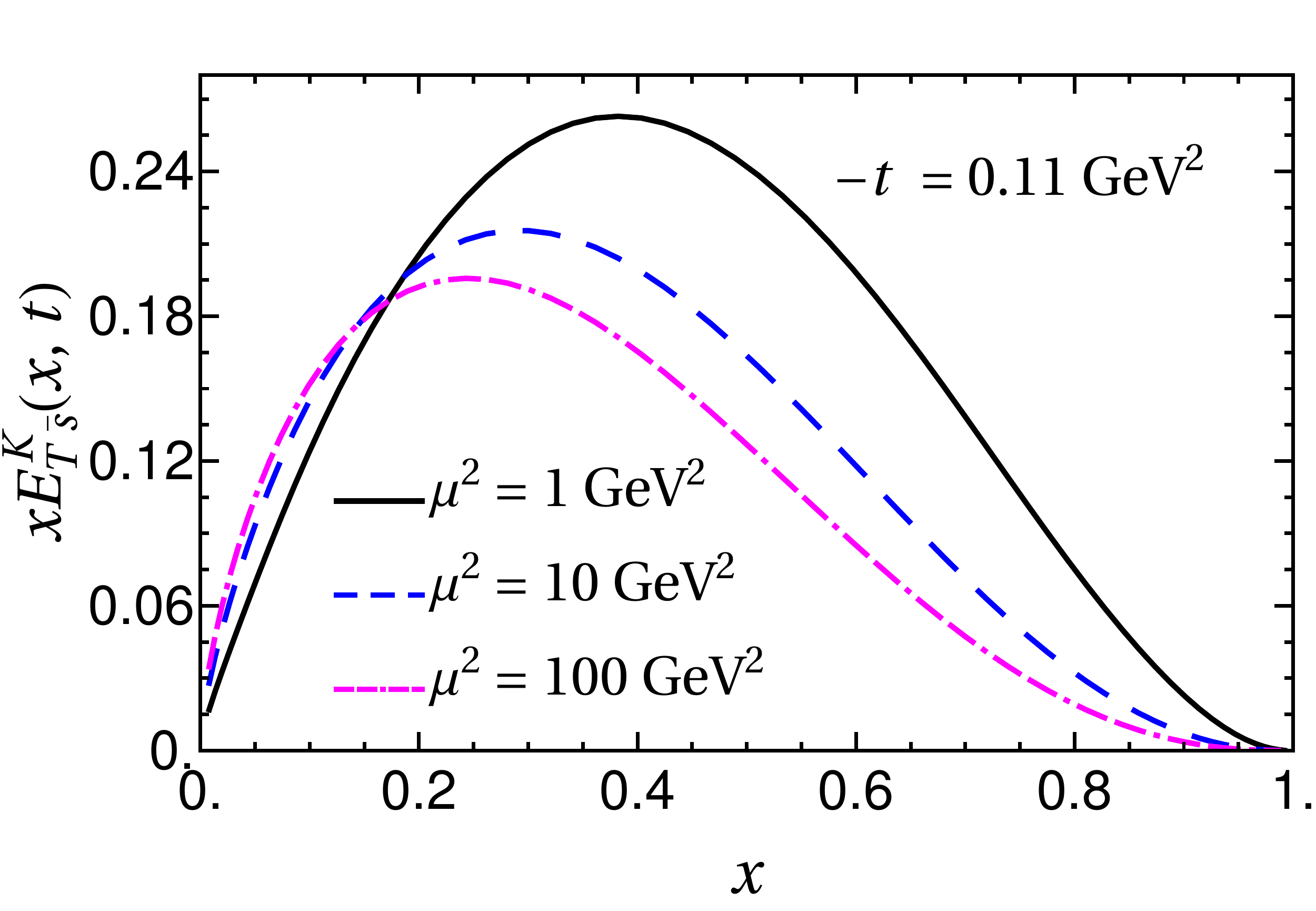}}
\end{tabular}
\caption{Scale evolution of the valence quark GPDs of the kaon: (a) $H(x,0,t)$ and (c) $E_T(x,0,t)$ are for the valence $u$ quark: (b) and (d) are same as (a) and (c), respectively, but for the valence $\bar{s}$ quark as functions of $x$ and fixed value of $-t=0.11$ GeV$^2$. The GPDs are evaluated  with $N_\text{max} = 8$ and $L_\text{max} = 32$ within the BLFQ-NJL model. The GPDs are evolved from our model scale for the kaon $\mu_{0K}^2=0.246$ GeV$^2$ to the final scales $\mu^2=1,~10,~100$ GeV$^2$.}
\label{evolution_kaon_gpds}
\end{figure*}

\begin{figure*}
\begin{tabular}{cc}
\subfloat[]{\includegraphics[scale=0.3]{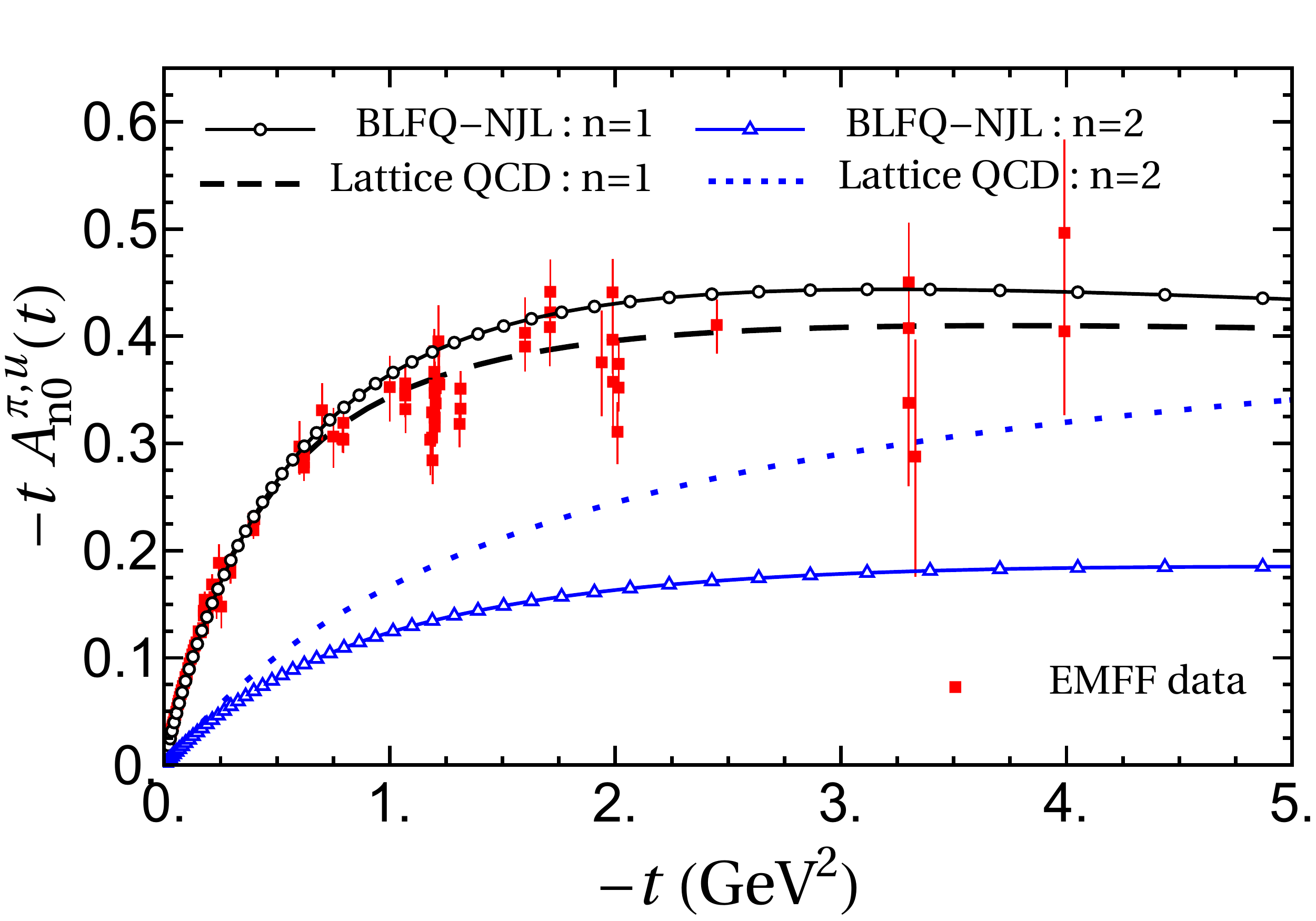}}
\end{tabular}
\begin{tabular}{cc}
\subfloat[]{\includegraphics[scale=0.3]{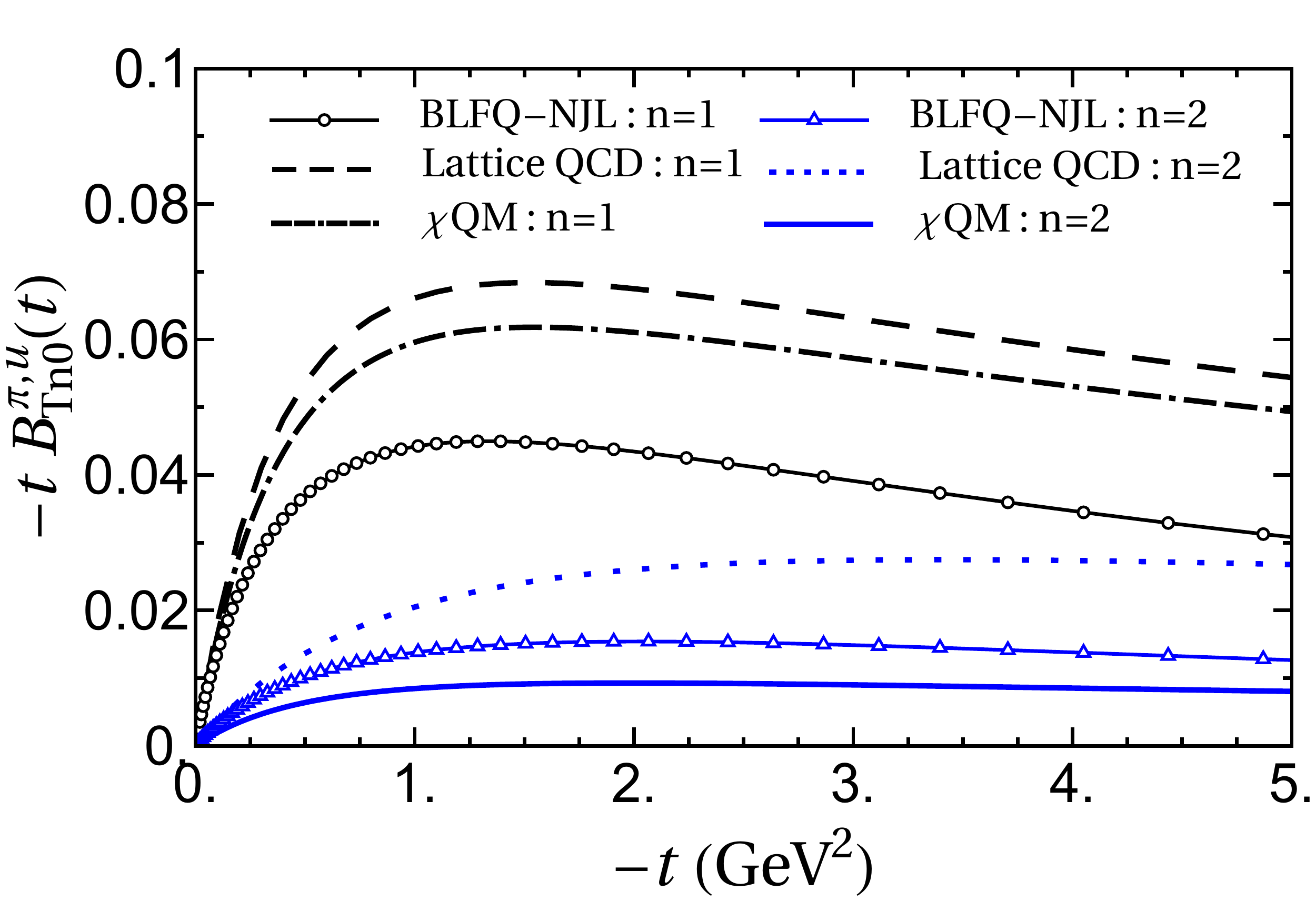}}
\end{tabular}
\caption{The first two Mellin moments of the valence quark GPDs of the pion: (a) $-tA^{\pi,u(\bar{d})}_{n0}(t)$ and (b) $-tB^{\pi,u(\bar{d})}_{Tn0}(t)$ for $n=1$ (black lines) and $n=2$ (blue lines) as functions of $-t$. 
The electromagnetic form factor $A_{10}(t)$ of the pion is compared with the experimental data~\cite{Amendolia:1986wj,Bebek:1974iz,Bebek:1974ww,Bebek:1977pe,Volmer:2000ek,Horn:2006tm} and the lattice QCD result~\cite{Brommel:2006ww}. The gravitational FF $A_{20}(t)$ is compared with the parameterization of lattice QCD simulations at $\mu^2=4$ GeV$^2$, while $B_{T10}(t)$ and $B_{T20}(t)$ are compared with lattice QCD and the $\chi$QM results at the same scale $\mu^2=4$ GeV$^2$. 
The lines with circle and triangle symbols correspond to the results calculated in the BLFQ-NJL model (present work). The dashed ($n=1$) and dotted ($n=2$) lines represent the lattice QCD results~\cite{Brommel:2007xd}, whereas the dash-dotted ($n=1$) and solid ($n=2$) lines in (b) represent the $\chi$QM~\cite{Nam:2010pt} results. The experimental results in (a) are for the EMFF only.
}
\label{pion_moments}
\end{figure*}
\begin{figure*}
\begin{tabular}{cc}
\subfloat[]{\includegraphics[scale=0.3]{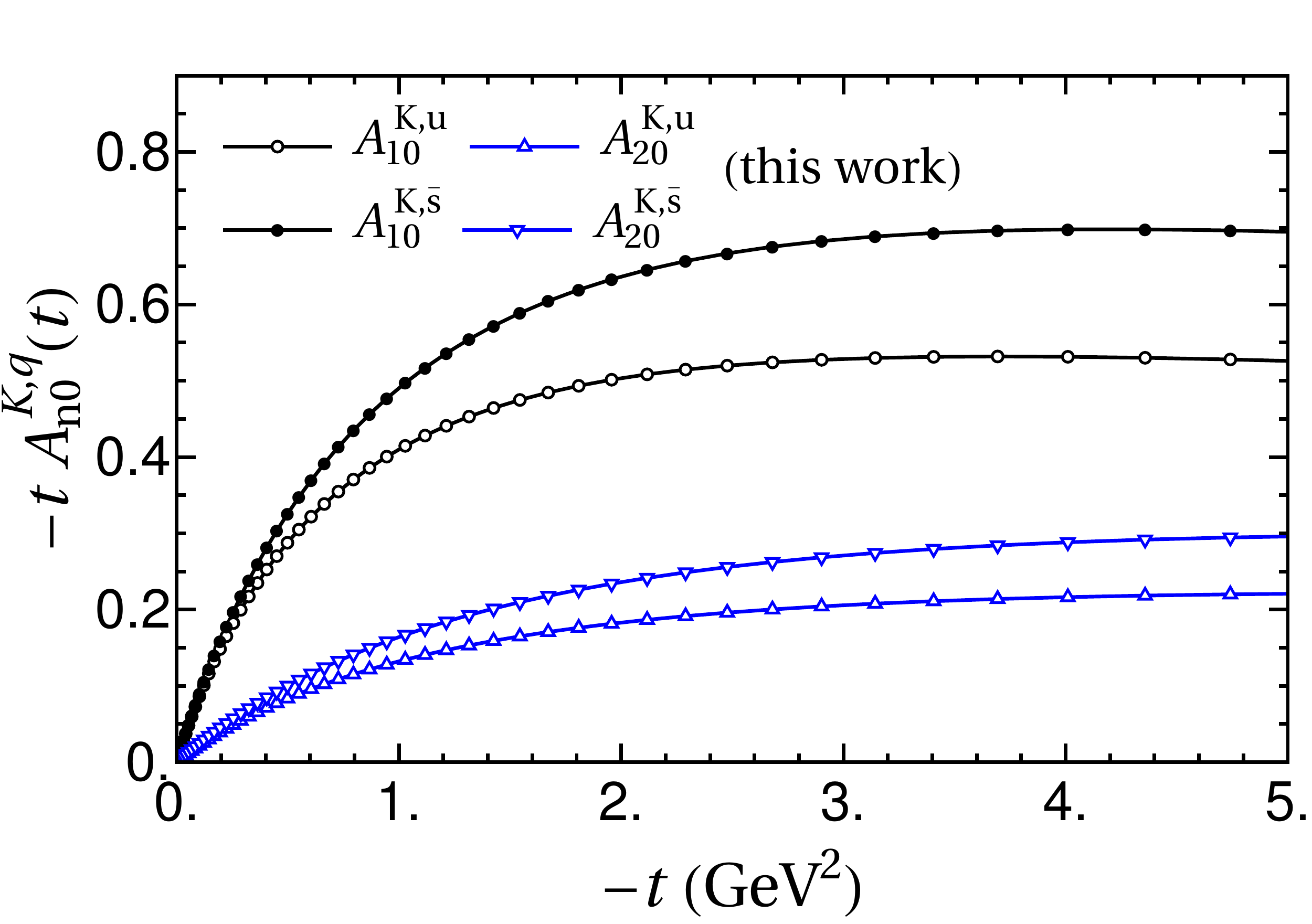}}
\end{tabular}
\begin{tabular}{cc}
\subfloat[]{\includegraphics[scale=0.15]{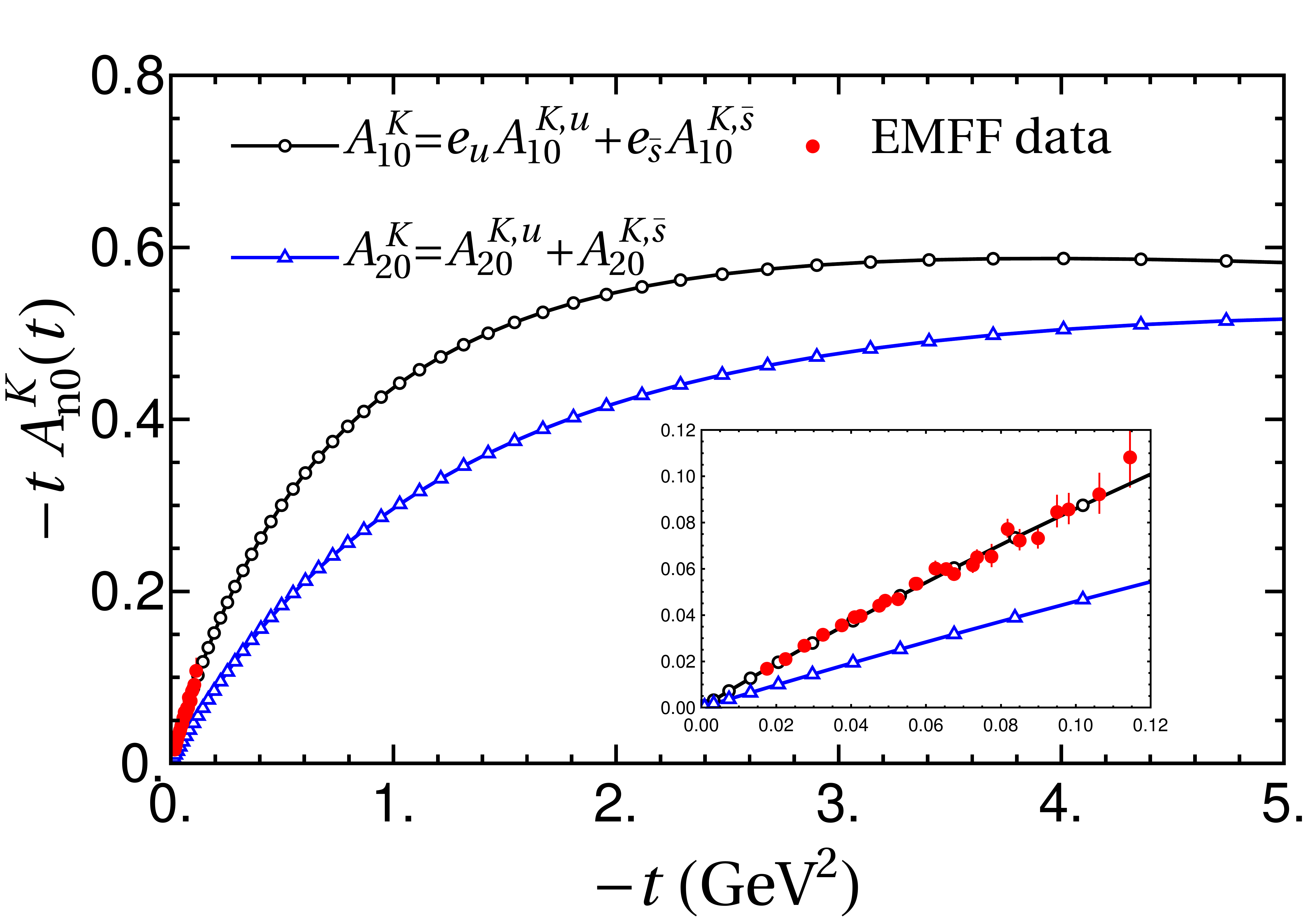}}\\
\end{tabular}
\begin{tabular}{cc}
\subfloat[]{\includegraphics[scale=0.335]{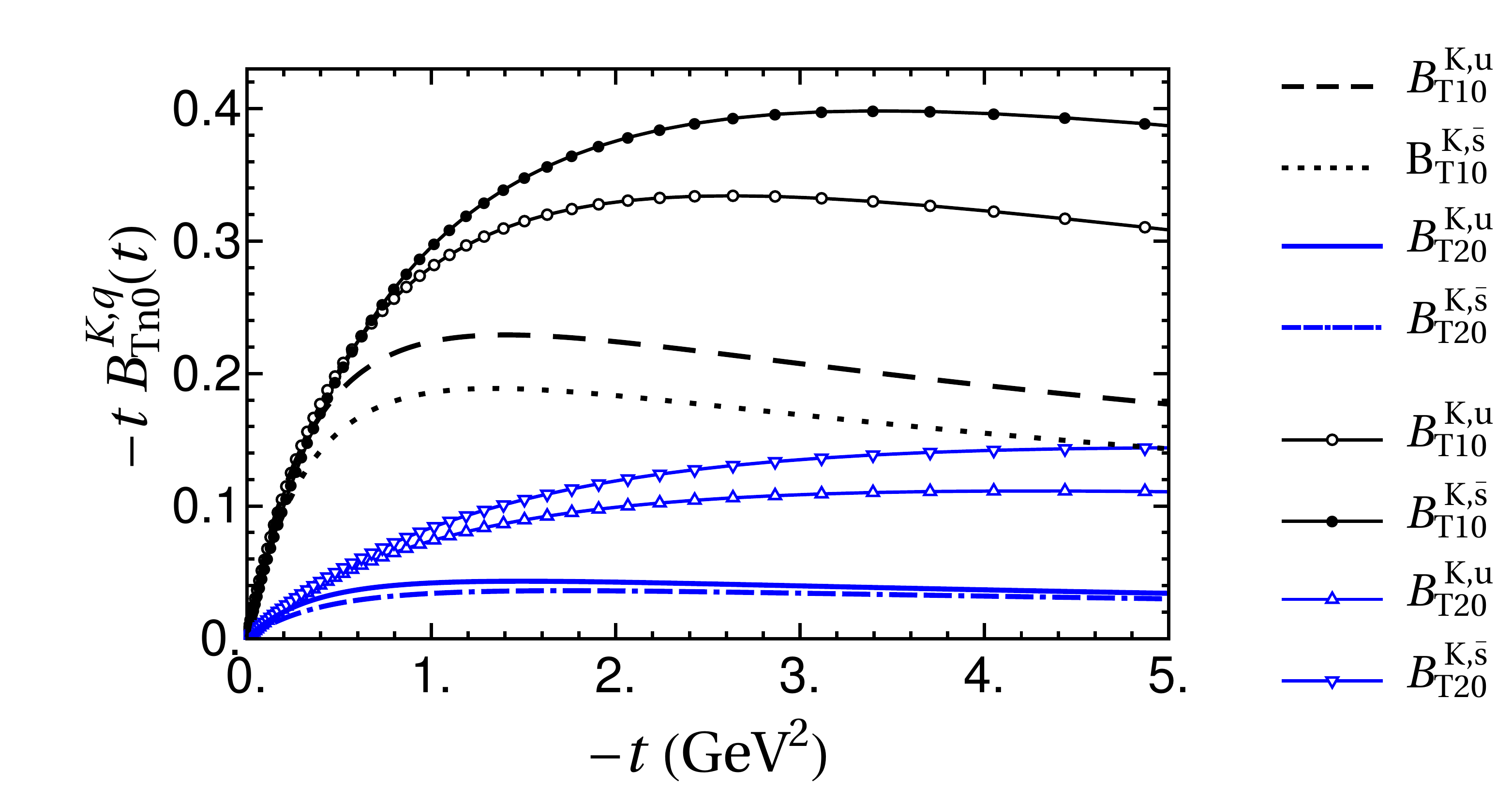}}
\end{tabular}
\caption{The first two Mellin moments of the valence quark GPDs of the kaon: (a) $-tA^{K,u(\bar{s})}_{n0}(t)$, (b) $-tA^{K}_{n0}(t)$, and (c) $-tB^{K,u(\bar{s})}_{Tn0}(t)$ for $n=1$ (black lines) and $n=2$ (blue lines) as functions of $-t$. 
The electromagnetic form factor $A_{10}(t)$ of the kaon in (b) is compared with the experimental data~\cite{Dally:1980dj,Amendolia:1986ui}. The inset in (b) provides an expanded view of the EMFF at low $-t$. The tensor FFs in (c) $B_{T10}(t)$ and $B_{T20}(t)$ are compared with the $\chi$QM~\cite{Nam:2011yw} at $\mu^2=4$ GeV$^2$. The lines with circle and triangle symbols correspond to the results calculated in the BLFQ-NJL model (this work). The lines without symbols in (c) represent the $\chi$QM results. The experimental results in (b) are for the EMFF only.  
}
\label{kaon_H_moments}
\end{figure*}
\section{numerical results and discussion \label{result}}
\subsection{GPDs and generalised form factors \label{FFs}}
The LFWFs of the valence quarks in the pion and the kaon~\cite{Jia:2018ary} have been solved in the BLFQ framework using the NJL interactions as briefly discussed in the Subsection~\ref{sc:BLFQ_NJL}. 
We insert the valence wave functions given by Eq.~\eqref{eq:psi_rs_basis_expansions} into Eqs.~\eqref{eq:H_valence_psi} and \eqref{eq:ET_valence_psi} to calculate the GPDs for the pion and the kaon. We employ the wave functions obtained at the basis truncation $N_\text{max} = 8$ and $L_\text{max} = 32$ with other model parameters given in Table~\ref{tab:model_parameters}.
We illustrate the valence GPDs $H^q$ and  $E_T^q$ ($q\equiv u$ or $\bar{d}$) as functions of $x$ and $-t$ for the pion in Fig.~\ref{pion_gpds}. In the forward limit ($-t=0$), the unpolarized GPD $H$ reduces to the ordinary PDF, which peaks at $x=0.5$ for the pion, reflecting the symmetry between the valence quark and the valence antiquark. Unlike the unpolarized GPD $H$, the chiral-odd GPD, $E_T$ in the pion has its peak located below the central value of $x$ and is asymmetric under $x\leftrightarrow (1-x)$ even when $-t=0$. This is due to the fact that $E_T$ involves the overlaps of the wavefunctions with different orbital angular momentum $L_z=0$ and $L_z=\pm 1$. The peaks of these GPDs shift towards higher values of $x$ and the magnitudes of distributions decrease with increasing the value of $-t$. 

The valence quark GPDs for the kaon are shown in Fig.~\ref{kaon_gpds}. The up quark GPD $H(x,0,t)$ in the kaon, unlike the valence quark GPD $H(x,0,t)$ in the pion,  has the maximum at lower $x$ ($<0.5$) when $t=0$, whereas, due to its heavy mass, the peak in the strange quark distribution appears at higher $x$ ($>0.5$). Meanwhile, the peaks along $x$ get shifted to larger values of $x$ with increasing $-t$ similar to that  observed in the pion GPD. This seems to be a model independent behavior of the GPDs which has been noticed in other phenomenological models for the pion~\cite{Kaur:2018ewq} as well as for the nucleon~\cite{Chakrabarti:2013gra,Mondal:2015uha,Chakrabarti:2015ama,Xu:2021wwj}. We also notice that the GPD $E_T$ for the up quark in the kaon exhibits a behavior similar to that  observed in the pion, however, the magnitude of $E_{ Tu}$ in the kaon is larger than that in the pion. Meanwhile, $E_{ Ts}$ displays a different behavior compared to $E_{Tu}$ in the kaon: $E_{ Ts}$ is broader along $x$ and falls slower at large $x$ compared to $E_{ Tu}$. As $-t$ increases, $E_{ Tu}$ also falls faster than $E_{ Ts}$ in the kaon. One can also observe oscillations in the GPDs along $x$ in Fig.~\ref{pion_gpds} and Fig.~\ref{kaon_gpds}, which are numerical artifacts due to longitudinal cutoff $L_{\rm max}$. The amplitudes of the oscillations decrease with increasing $L_{\rm max}$~\cite{Lan:2019rba}.

By performing the QCD evolution, the valence quark GPDs at high $\mu^2$ scale can be obtained with the input GPDs at the model scale $\mu_0^2$. 
We adopt the DGLAP equations \cite{Dokshitzer:1977sg,Gribov:1972ri,Altarelli:1977zs}  of QCD with NNLO for this scale evolution.
Explicitly, we evolve our input GPDs to the relevant experimental scales with independently adjustable initial scales of the pion and the kaon GPDs utilizing the higher order perturbative parton evolution toolkit (HOPPET)~\cite{Salam:2008qg}. We adopt ${\mu_{0\pi}^2=0.240\pm0.024~\rm{GeV}^2}$ for the initial scale of the pion GPDs and ${\mu_{0K}^2=0.246\pm0.024~\rm{GeV}^2}$ for the initial scale of the kaon GPDs which we determined by requiring the results after NNLO DGLAP evolution to fit both the pion and the kaon PDFs results from the experiments~\cite{Lan:2019rba}.
We show the valence quark GPDs in the pion and the kaon for a fixed value of $-t$ at different $\mu^2$ evolved from the corresponding initial scales in Fig.~\ref{evolution_pion_gpds} and Fig.~\ref{evolution_kaon_gpds}, respectively. We observe that the peaks of the distributions move to lower $x$ as we evolve the GPDs to higher scales. The moments of the distributions decrease uniformly as the scale $\mu^2$ increases. 
 The qualitative behavior of the evolved GPDs is similar in both the pion and the kaon.

The Mellin moments of the valence GPDs give the generalized FFs. The Mellin moments  are defined as~\cite{Brommel:2007xd}
\be
A^q_{n0}(t)=\int_0^1~dx\, x^{n-1}\,H^q(x,0,t),\label{moment_formula_A}\\
B^q_{Tn0}(t)=\int_0^1~dx\, x^{n-1}\,E_T^q(x,0,t),\label{moment_formula_B}
\ee
where the index $n=1,2,3\dots$, and the second subscript corresponds to the fact that the moments are evaluated at zero skewness ($\zeta=0$). 
The first moments of the unpolarized GPD ${H}^q(x,0,t)$ give the electromagnetic FF, $F^q(t)=A^q_{10}(t)$ of an unpolarized quark, while in the forward limit, i.e, $t = 0$, the FF $F^q(0)$ gives the number of the valence quarks of flavor $q$.  
The first moment of chiral-odd GPD ${E}^q_{T}(x,0,t)$ provides the tensor FF $B_T^q(t)$ when the quark is transversely polarized. 
The second moments of these GPDs correspond to the gravitational FFs of the quarks. Meanwhile, the third moments of the GPDs provide the FFs of a twist-two operator having two covariant derivatives~\cite{Diehl:2003ny,Belitsky:2005qn} and the higher order moments produce the FFs of higher-twist operators.

\begin{table*}
\caption{BLFQ-NJL model predictions for $A^{\pi(K),q}_{20}(0)$, $B^{\pi(K),q}_{T10}(0)$ and $B^{\pi(K),q}_{T20}(0)$ at the scale 4 GeV$^2$. We compare our results  with the available lattice QCD simulations~\cite{Brommel:2007xd},
the $\chi$QMs~\cite{Nam:2010pt,Nam:2011yw}, and the CCQM~\cite{Fanelli:2016aqc} at the same scale 4 GeV$^2$. The errors in our results correspond to the QCD evolution from the initial scales ${\mu_{0\pi}^2=0.240\pm0.024~\rm{GeV}^2}$ for the pion and ${\mu_{0K}^2=0.246\pm0.024~\rm{GeV}^2}$ for the kaon.} \label{tab:tensor_charge}
 \centering
\begin{tabular}{ccc ccc ccc ccc ccc ccc ccc}
\toprule
 Quantity ~&~ BLFQ-NJL  ~&~  Lattice QCD~\cite{Brommel:2007xd}   ~&~  $\chi$QM~\cite{Nam:2010pt}  ~&~ $\chi$QM~\cite{Nam:2011yw}   ~&~ $\chi$QM~\cite{Nam:2011yw} ~&~ CCQM~\cite{Fanelli:2016aqc} ~&\\
   ~&~ (this work)  ~&~   ~&~    ~&~ (model I)  ~&~ (model II)& &\\
\hline
\vspace{0.1cm}
$A^{\pi,q}_{20}(0)$ & $0.244\pm 0.018$ & $0.27 \pm 0.01$ & ... & ... & ... & $0.248$ & \\
\vspace{0.1cm}
$A^{K,u}_{20}(0)$ & $0.235 \pm 0.018$ & ... & ... & ... & ... & ... & \\
\vspace{0.1cm}
$A^{K,\bar{s}}_{20}(0)$ & $0.265 \pm 0.020$ & ... & ... & ... & ... & ...& \\
\hline
\vspace{0.1cm}
$B^{\pi,q}_{T10}(0)$ & $0.229\pm 0.004$ & $0.216 \pm 0.034$ & $0.216$ & ... & ...  & $0.126$ & \\\vspace{0.1cm}
$B^{K,u}_{T10}(0)$ & $0.821 \pm 0.014$ & ... & ... & $0.783$ & $0.611$ & ... & \\
\vspace{0.1cm}
$B^{K,\bar{s}}_{T10}(0)$ & $0.706\pm0.010$ & ... & ... & $0.676$ & $0.421$ & ... & \\
\hline
\vspace{0.1cm}
$B^{\pi,q}_{T20}(0)$ & $0.045\pm0.004$ & $0.039 \pm 0.010$ & $0.032$ & ... & ... & $0.028$ &\\
\vspace{0.1cm}
$B^{K,u}_{T20}(0)$ & $0.152\pm 0.011$ & ... & ... & $0.139$ & $0.090$ & ... & \\
\vspace{0.1cm}
$B^{K,\bar{s}}_{T20}(0)$ & $0.152\pm 0.011$ & ... & ... & $0.100$ & $0.076$ & ... & \\
\botrule
\end{tabular}
\end{table*}
\begin{figure*}
\begin{tabular}{cc}
\subfloat[]{\includegraphics[scale=0.45]{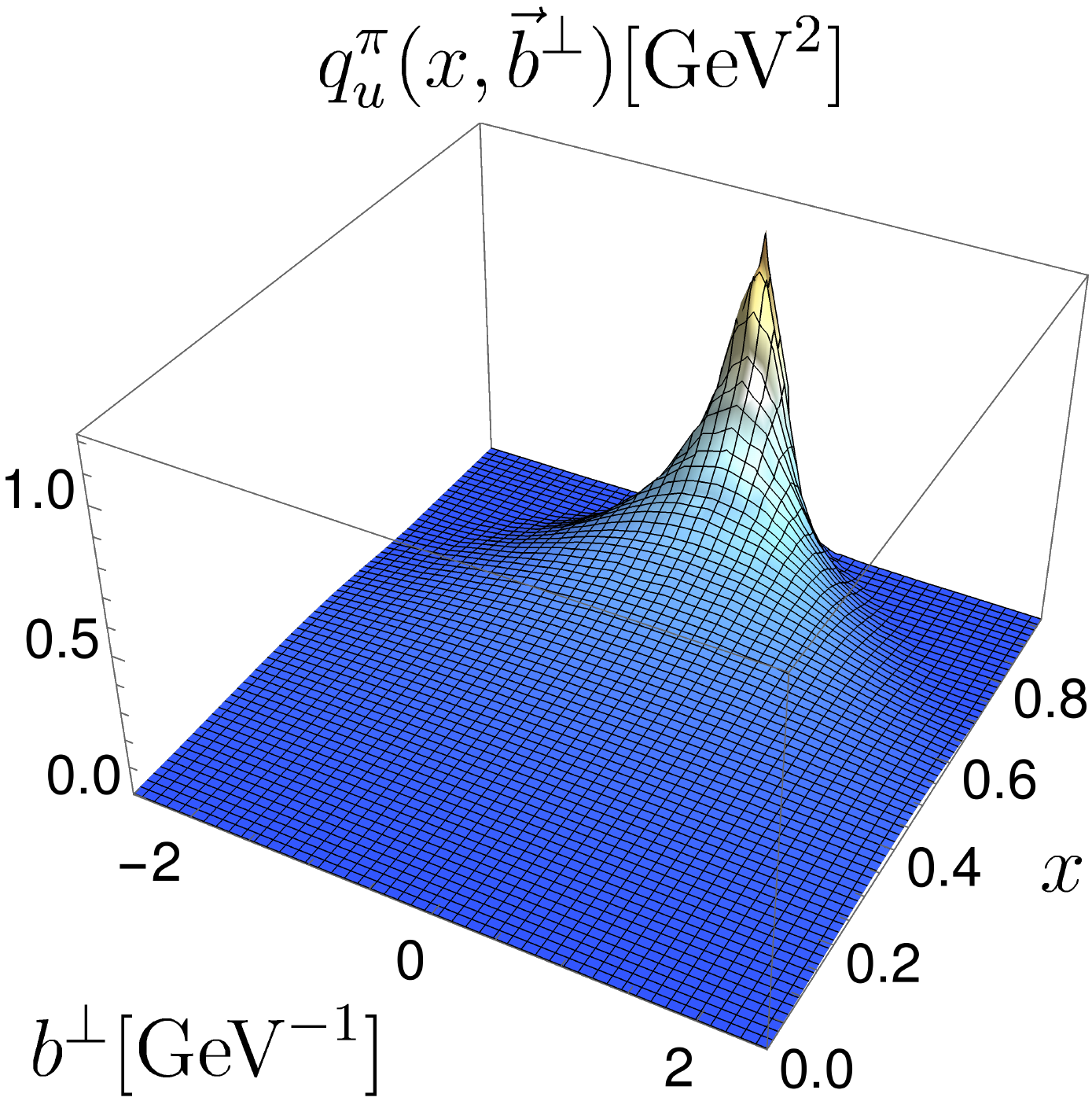}}
\end{tabular}
\begin{tabular}{cc}
\subfloat[]{\includegraphics[scale=0.45]{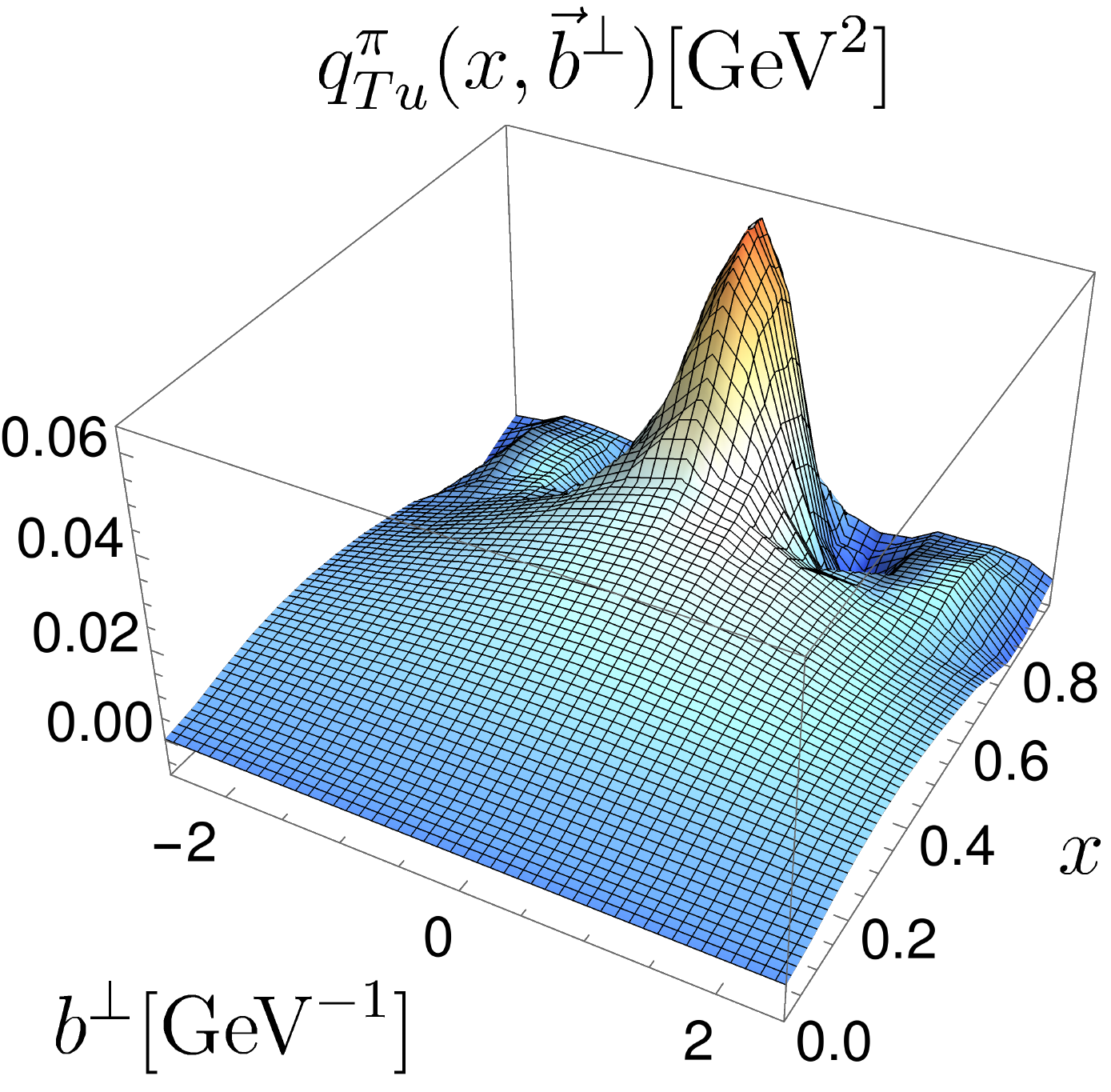}}
\end{tabular}
\caption{The valence ($u$ or $\bar{d}$) quark GPDs for the pion in the transverse impact parameter space: (a) $q(x, {\vec b}^{\perp})$ and (b) $q_T(x,{\vec b}^{\perp})$ as functions of $x$ and $b^\perp$, 
where $b^\perp$ is the transverse distance of the active quark from the center of momentum  of the hadron.}
\label{pion_impact_gpds}
\end{figure*}
\begin{figure*}
\begin{tabular}{cc}
\subfloat[]{\includegraphics[scale=0.45]{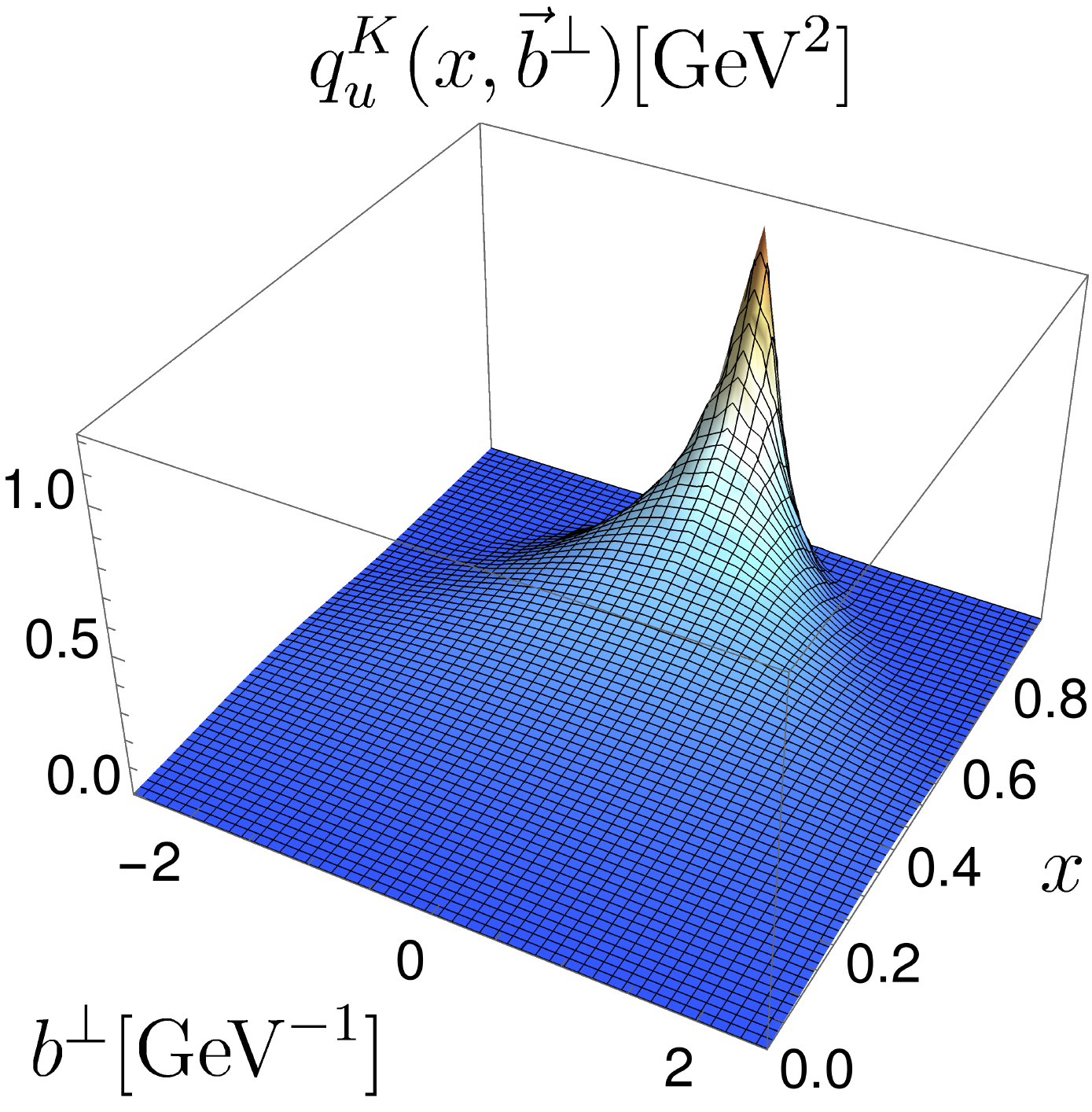}}
\end{tabular}
\begin{tabular}{cc}
\subfloat[]{\includegraphics[scale=0.45]{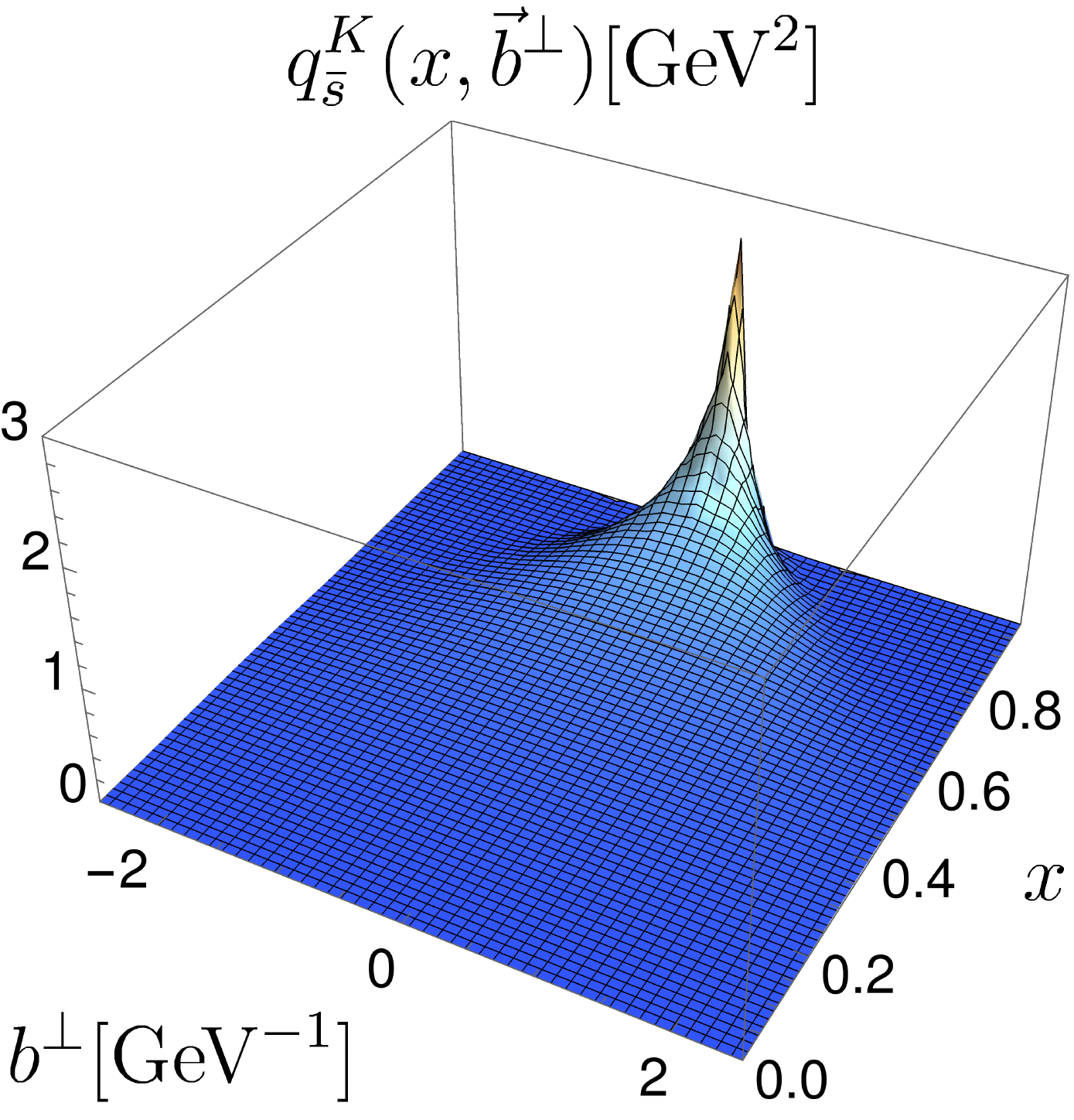}}
\end{tabular}
\begin{tabular}{cc}
\subfloat[]{\includegraphics[scale=0.45]{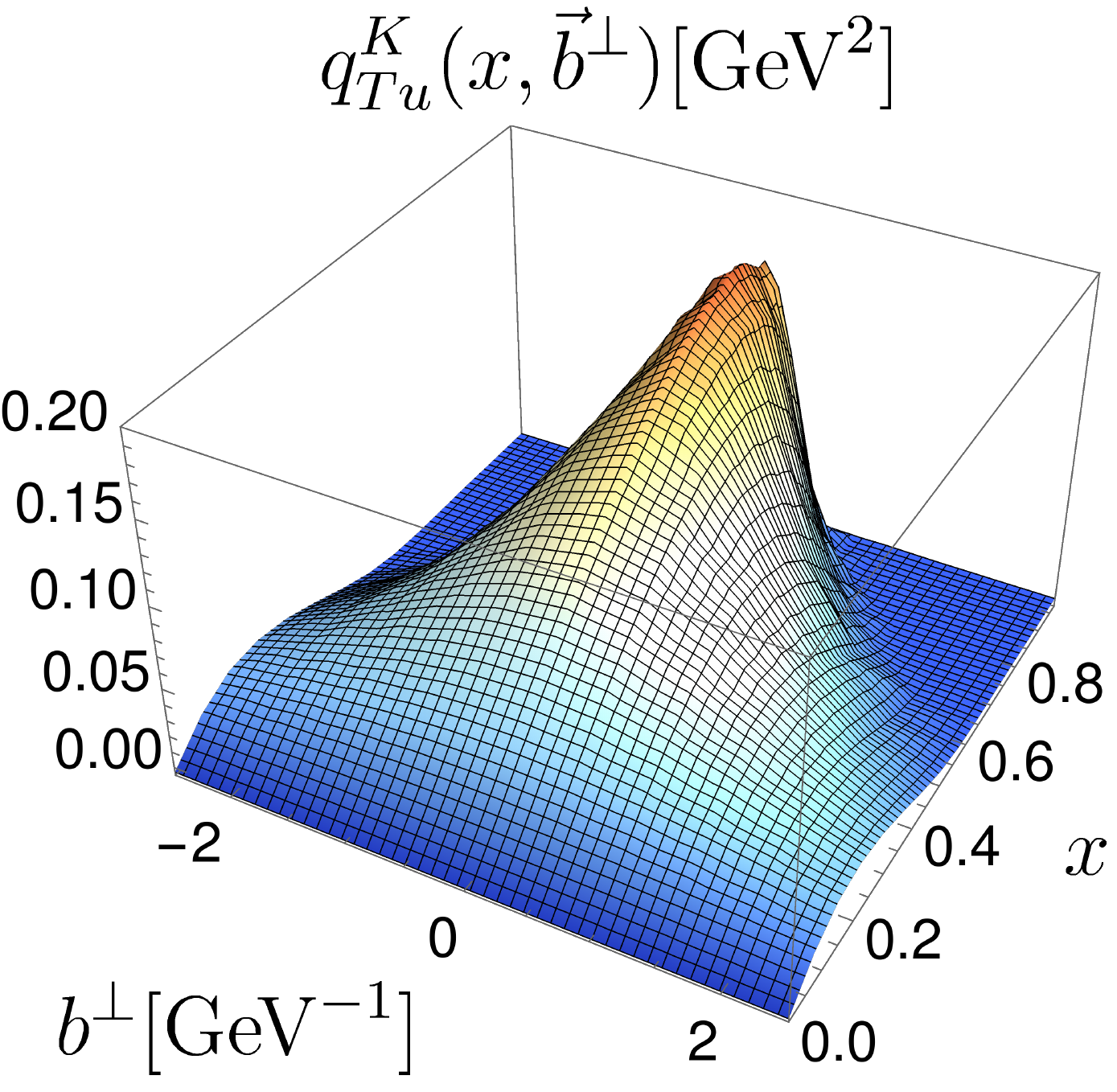}}
\end{tabular}
\begin{tabular}{cc}
\subfloat[]{\includegraphics[scale=0.45]{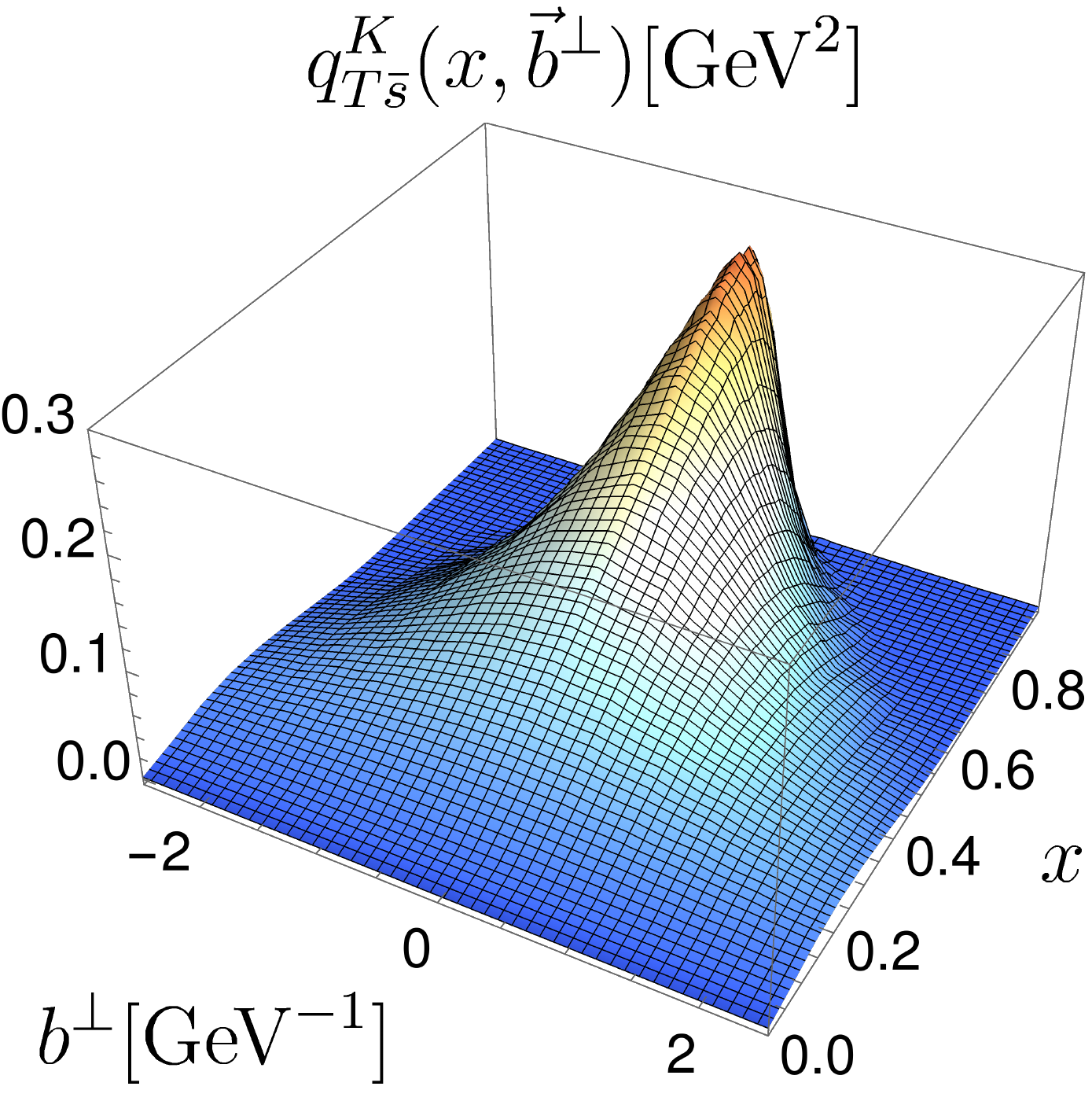}}
\end{tabular}
\caption{The valence quark GPDs of the kaon in the transverse impact parameter space: (a) $q(x, {\vec b}^{\perp})$ and (c) $q_T(x, {\vec b}^{\perp})$ are for the valence $u$ quark; (b) and (d) are same as (a) and (c), respectively, but for the valence $\bar{s}$ quark as functions of $x$ and $b^\perp$, where $b^\perp$ is the transverse distance of the active quark from the center of momentum  of the hadron.}
\label{kaon_impact_gpds}
\end{figure*}

In Fig.~\ref{pion_moments}(a), we present the first two moments of the GPD ${H}^q(x,0,t)$ of the pion. The EMFF of the pion is given by  $F^\pi(t)=e_u A^{\pi,u}_{10}(t)+e_{\bar{d}}A^{\pi,\bar{d}}_{10}(t)$, where $e_q$ denotes the charge of the quark $q$. We find that the pion EMFF within our BLFQ-NJL model is in good agreement with the experimental data and with the lattice QCD simulations. The second moment of the GPD ${H}^q(x,0,t)$ is the gravitational FF $A^q_{20}(t)$, which, at $t=0$, provides the momentum, $\langle x\rangle_q$ carried by the quark. For the pion $A^{\pi,u}_{20}(0)=A^{\pi,\bar{d}}_{20}(0)=0.5$ at the model scale. To compare with lattice QCD, we evolve the GPD to the relevant scale. As summarized in Table~\ref{tab:tensor_charge}, we obtain that at $\mu^2=4$ GeV$^2$, $A^{\pi,q}_{20}(0)=0.244 \pm 0.018$, which is compatible with the result from the covariant constituent quark model (CCQM) model~~\cite{Fanelli:2016aqc}, while the lattice QCD provides the value of $0.27\pm 0.01$~\cite{Brommel:2007xd}. 
In addition, substantial difference between our BLFQ-NJL model and lattice QCD for $A^{\pi,q}_{20}(t)$ is observed when $-t$ is nonzero with disagreement increasing as $-t$ increases, as can be seen in Fig.~\ref{pion_moments}(a).  We show the tensor FFs of the pion in Fig.~\ref{pion_moments}(b), where we also compare the FFs $B^{\pi,q}_{T10}(t)$ and $B^{\pi,q}_{T20}(t)$ with the lattice QCD results evaluated at the physical pion mass~\cite{Brommel:2007xd}. At $\mu^2=4$ GeV$^2$, we obtain: $B_{T10}^{\pi,q}(0)=0.229\pm 0.004$ and $B_{T20}^{\pi,q}(0)=0.045\pm 0.004$, which reasonably agree with the lattice QCD simulations within the uncertainty: $B_{T10}^{\pi,q}(0)=0.216\pm0.034$ and $B_{T20}^{\pi,q}(0)=0.039\pm0.010$, respectively. It is notable that $B_{T10}^{\pi,q}(0)$ in the CCQM~\cite{Fanelli:2016aqc} differs significantly from our result.  The qualitative behavior of the tensor FFs $B^{\pi,q}_{T10}(t)$ and $B^{\pi,q}_{T20}(t)$ is also found to be comparable with the lattice QCD calculations and the chiral quark model ($\chi$QM)~\cite{Nam:2010pt} as shown in Fig.~\ref{pion_moments}(b).

Fig.~\ref{kaon_H_moments} shows the moments of the kaon GPDs. As can be seen from Fig.~\ref{kaon_H_moments}(a), the magnitude of  $-tA^{K,u}_{n0}(t)$ is lower than that for $\bar{s}$ quark, implying the faster fall-off of the $u$ quark EMFF compared to the $\bar{s}$ quark in the kaon as $-t$ increases. The EMFF of the kaon, $F^K(t)=e_u A^{K,u}_{10}(t)+e_{\bar{s}}A^{K,\bar{s}}_{10}(t)$, is in good agreement with the experimental data as shown in Fig.~\ref{kaon_H_moments}(b). This is expected because model parameters of the BLFQ-NJL model are partially determined based on the experimental charge radii. On the other hand, we obtain $A^{K,u}_{20}(0)=0.43$ and $A^{K,\bar s}_{20}(0)=0.57$ at the model scale, whereas at $\mu^2=4$ GeV$^2$, the corresponding values are $A^{K,u}_{20}(0)=0.235\pm 0.018$ and $A^{K,\bar s}_{20}(0)=0.265\pm 0.020$ as summarized in Table~\ref{tab:tensor_charge}. We also illustrate the $t$ dependence of the kaon gravitational FFs $A^{K,u}_{20}(t)$ and $A^{K,\bar s}_{20}(t)$  at $\mu^2=4$ GeV$^2$ in Fig.~\ref{kaon_H_moments}(a). The tensor FFs for the kaon in our BLFQ-NJL model are presented in Fig.~\ref{kaon_H_moments}(c), in comparison  with that of the $\chi$QM calculations (model I in Ref.~\cite{Nam:2011yw}). The qualitative behavior of $-tB^{K,q}_{Tn0}(t)$ in those models agree. At large $-t$, $-tB_{Tn0}(t)$ for the $\bar{s}$ quark is larger than that for the $u$ quark in the BLFQ-NJL model, while in the $\chi$QM one observes the opposite. We also compare the quark tensor FFs at $t=0$ in the kaon with the $\chi$QM in Table~\ref{tab:tensor_charge}.

\begin{figure*}
\begin{tabular}{cc}
\subfloat[]{\includegraphics[scale=0.25]{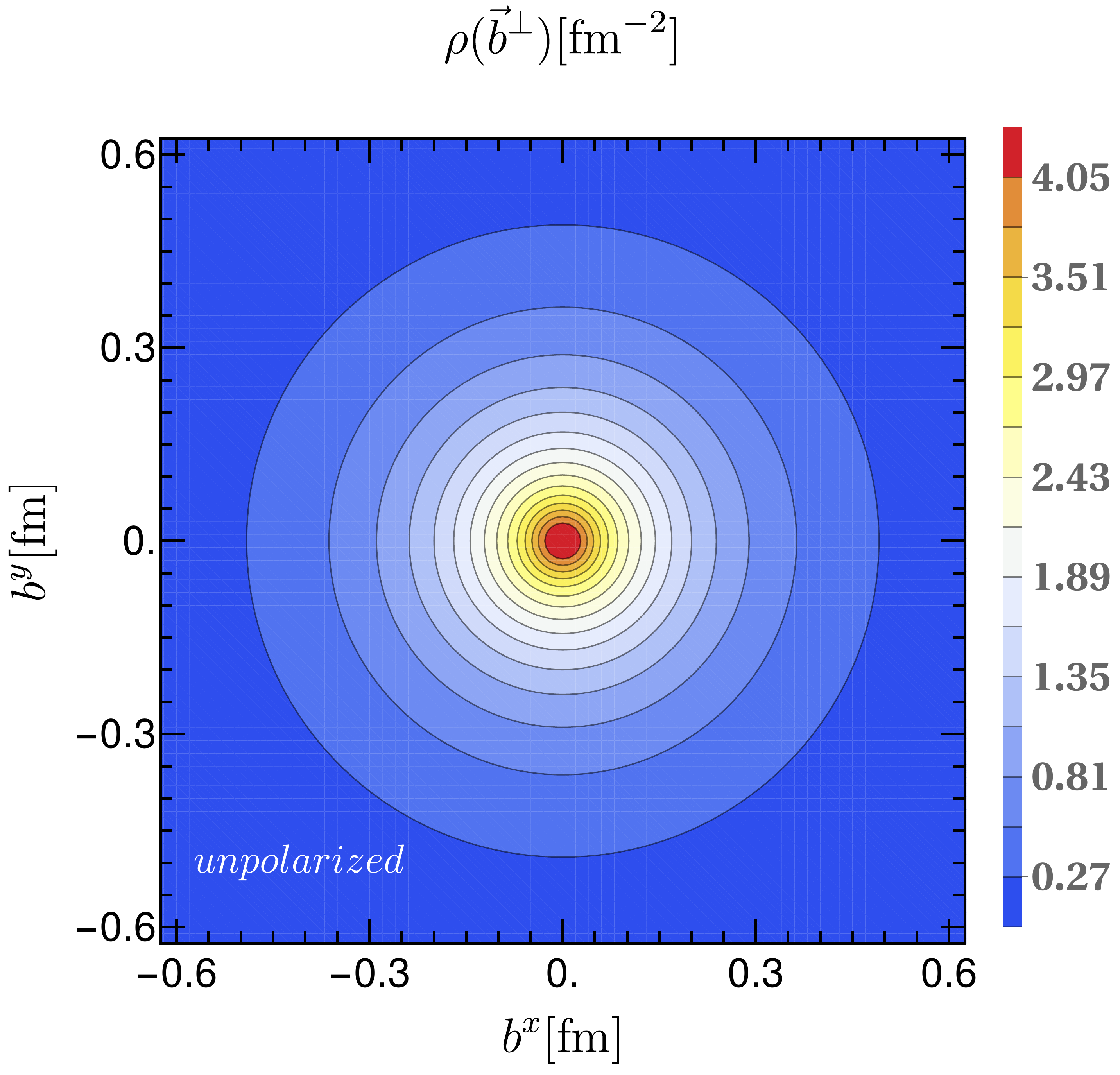}}
\end{tabular}
\begin{tabular}{cc}
\subfloat[]{\includegraphics[scale=0.25]{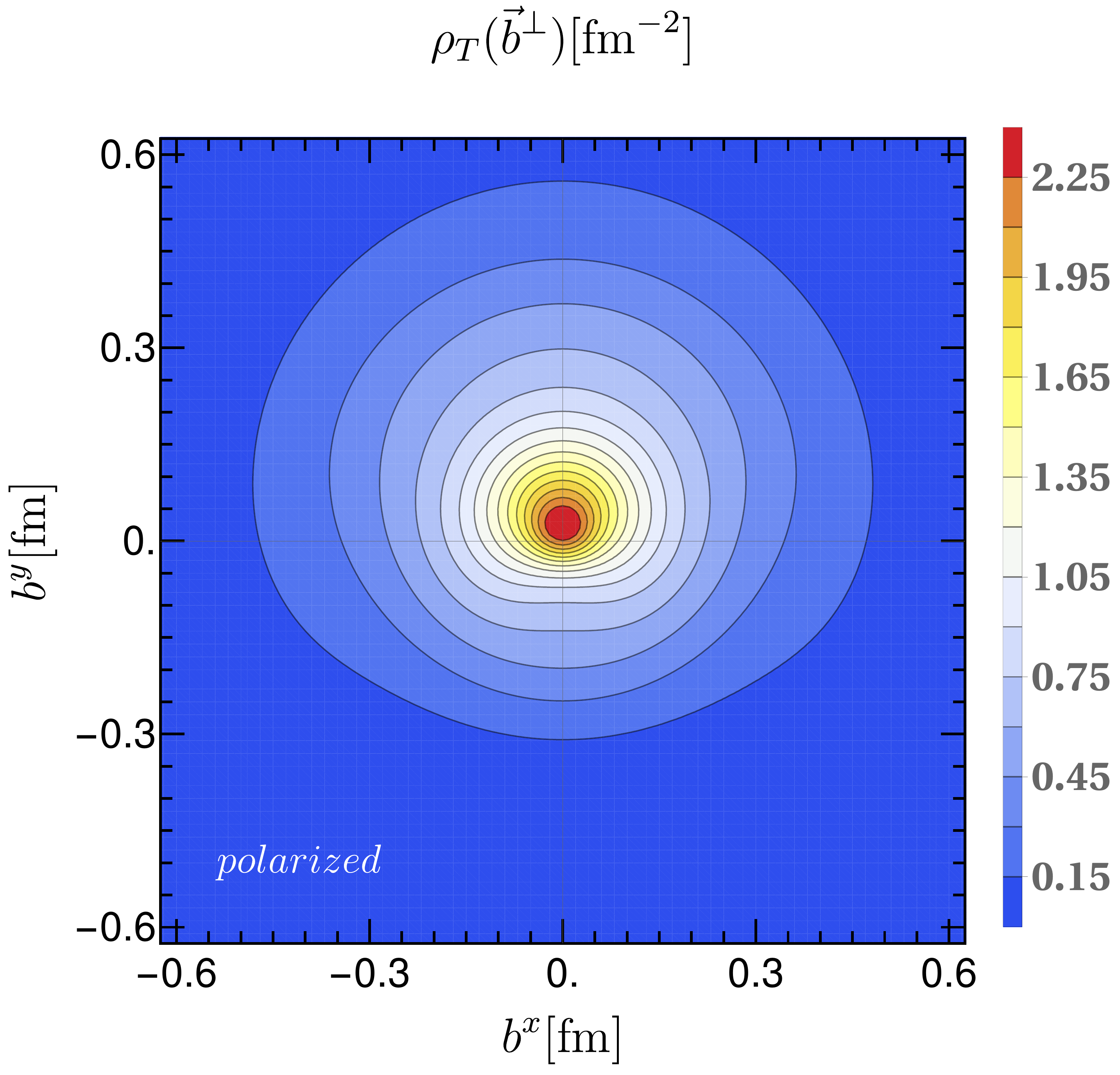}}
\end{tabular}
\begin{tabular}{cc}
\subfloat[]{\includegraphics[scale=0.3]{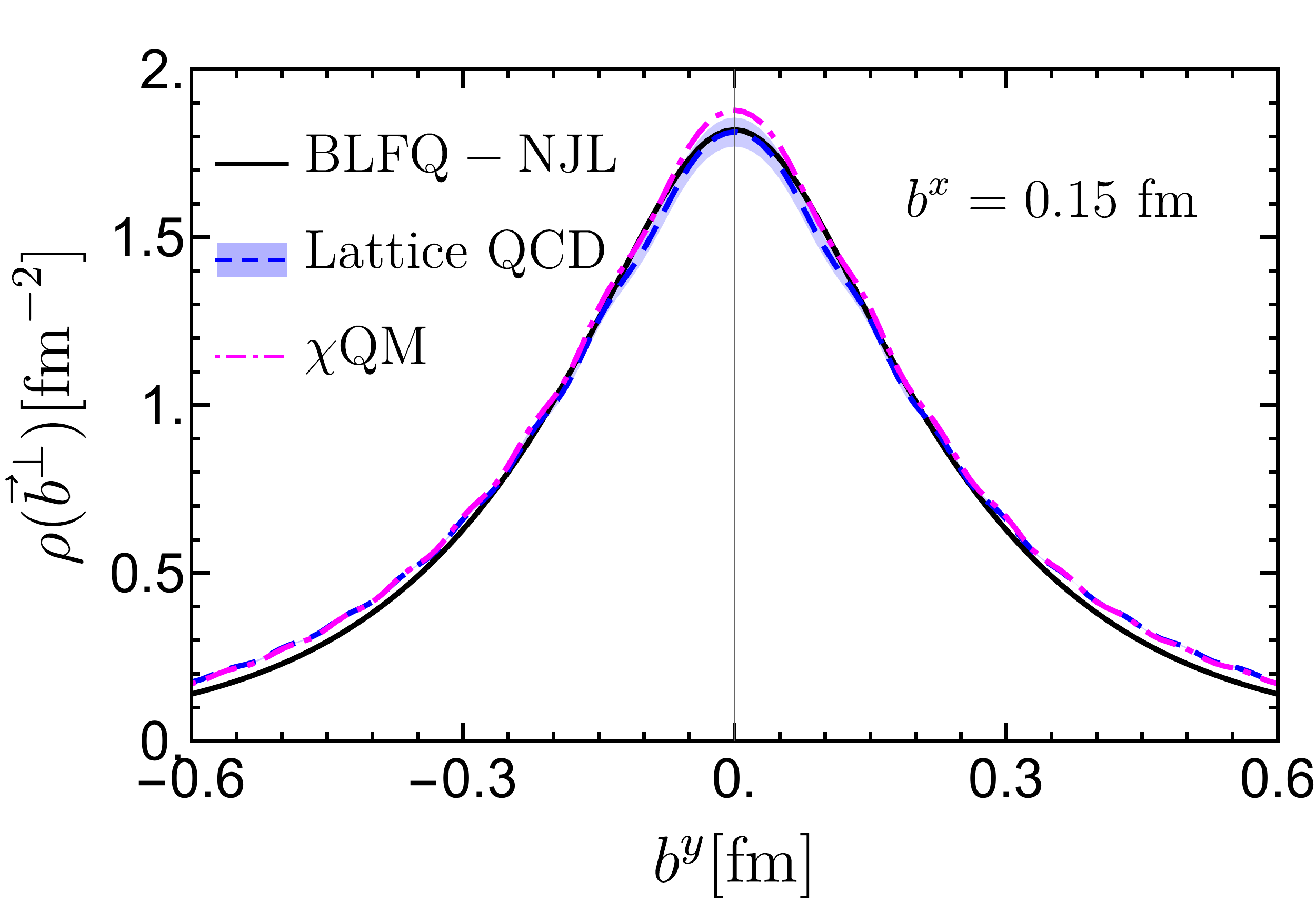}}
\end{tabular}
\begin{tabular}{cc}
\subfloat[]{\includegraphics[scale=0.3]{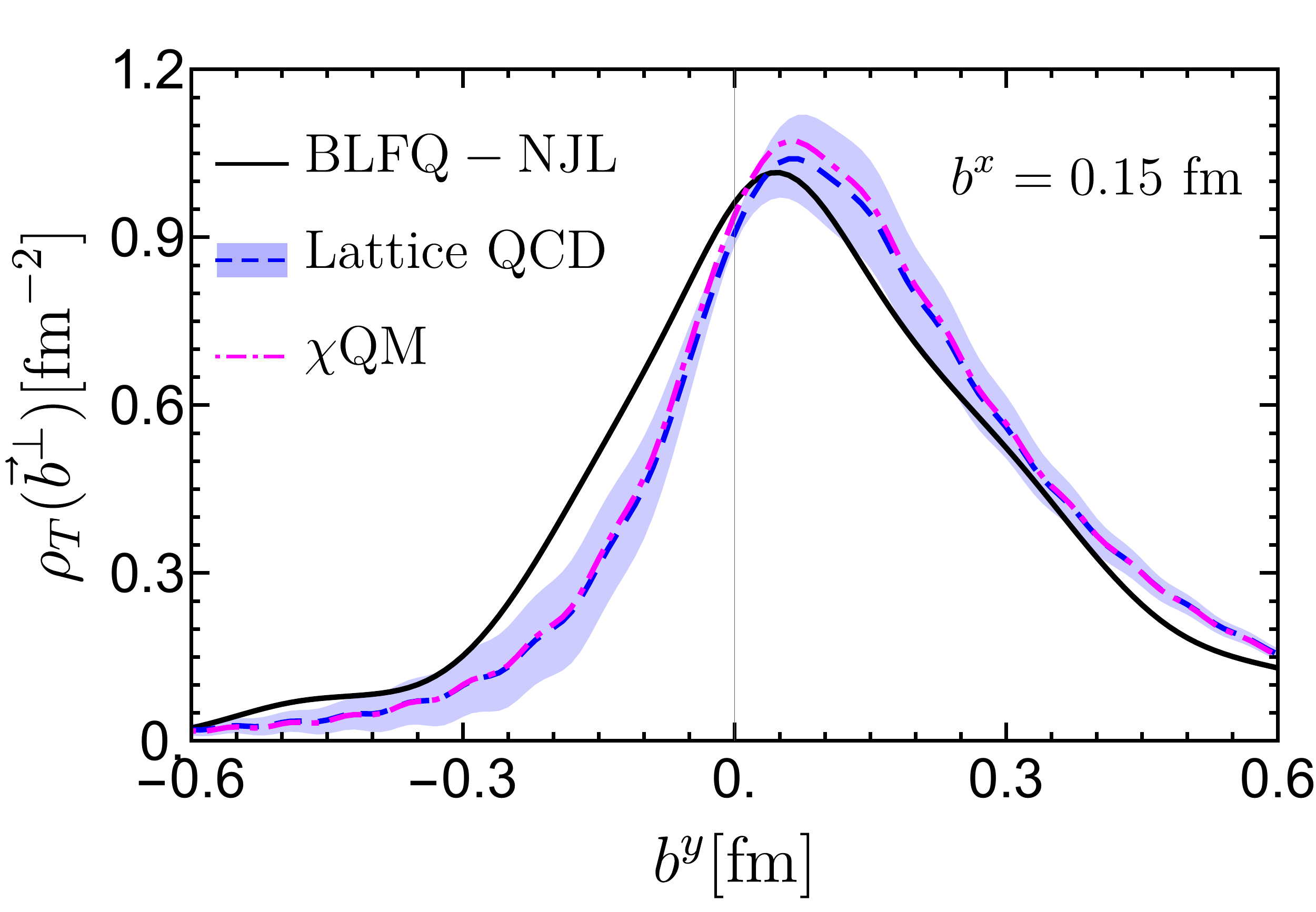}}
\end{tabular}
\caption{The valence quark probability density in the pion in the transverse impact parameter plane; (a) when the quark is unpolarized $\rho({\vec b}^{\perp})=\rho^{n=1}({\vec b}^{\perp},{\vec s}^{\perp}={\vec 0})$ and (b) when quark is transversely polarized along-$x$ direction $\rho_T({\vec b}^{\perp})=\rho^{n=1}({\vec b}^{\perp},{\vec s}^{\perp}=(1,0))$. Our corresponding results are compared with lattice QCD and the $\chi$QM model in (c) and (d), respectively. The solid, dotted and dash-dotted lines correspond to the BLFQ-NJL model, lattice QCD~\cite{Brommel:2007xd} and the $\chi$QM~\cite{Nam:2010pt}, respectively.}
\label{pion_density}
\end{figure*}
\begin{figure*}
\begin{tabular}{cc}
\subfloat[]{\includegraphics[scale=0.25]{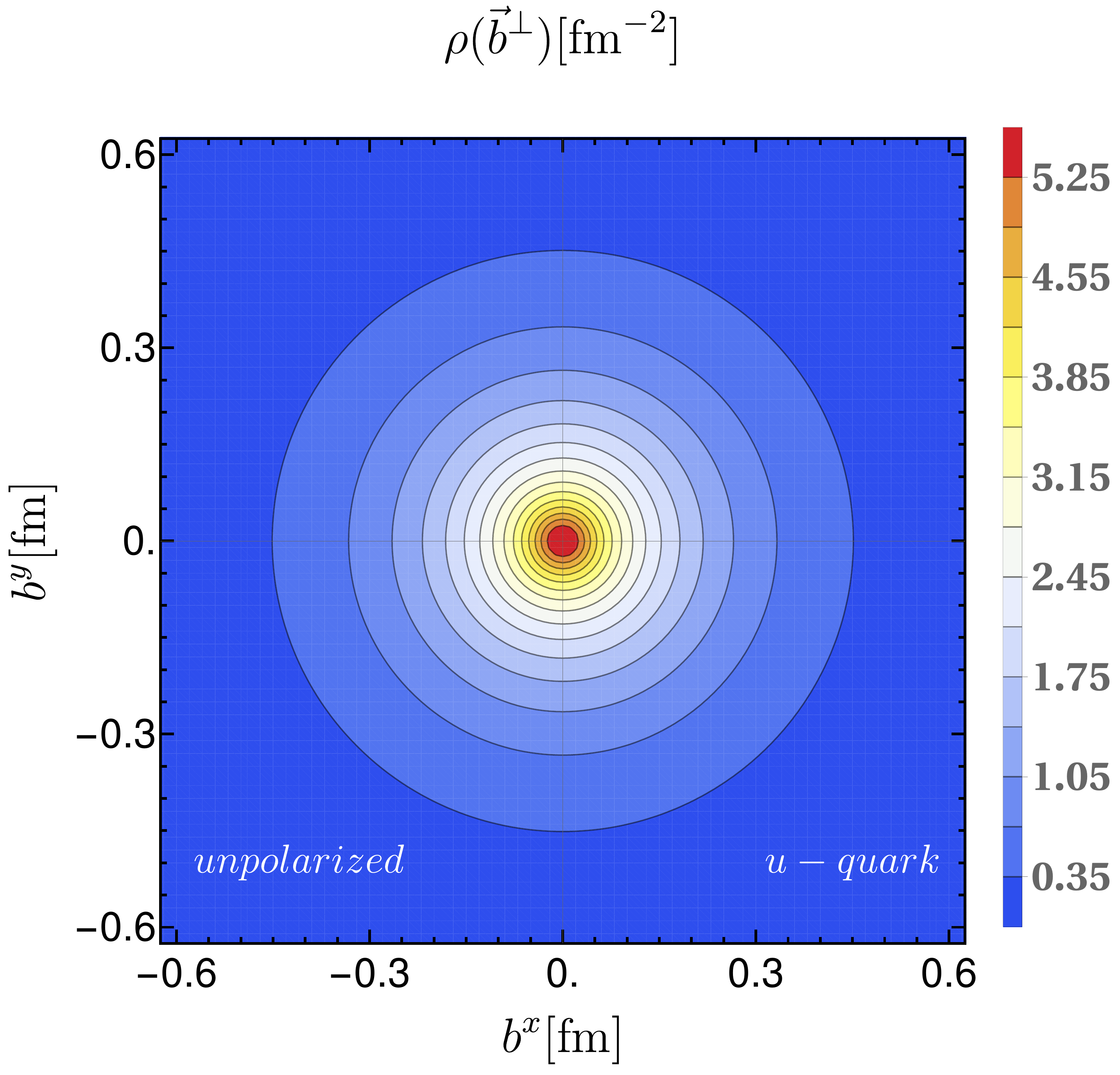}}
\end{tabular}
\begin{tabular}{cc}
\subfloat[]{\includegraphics[scale=0.25]{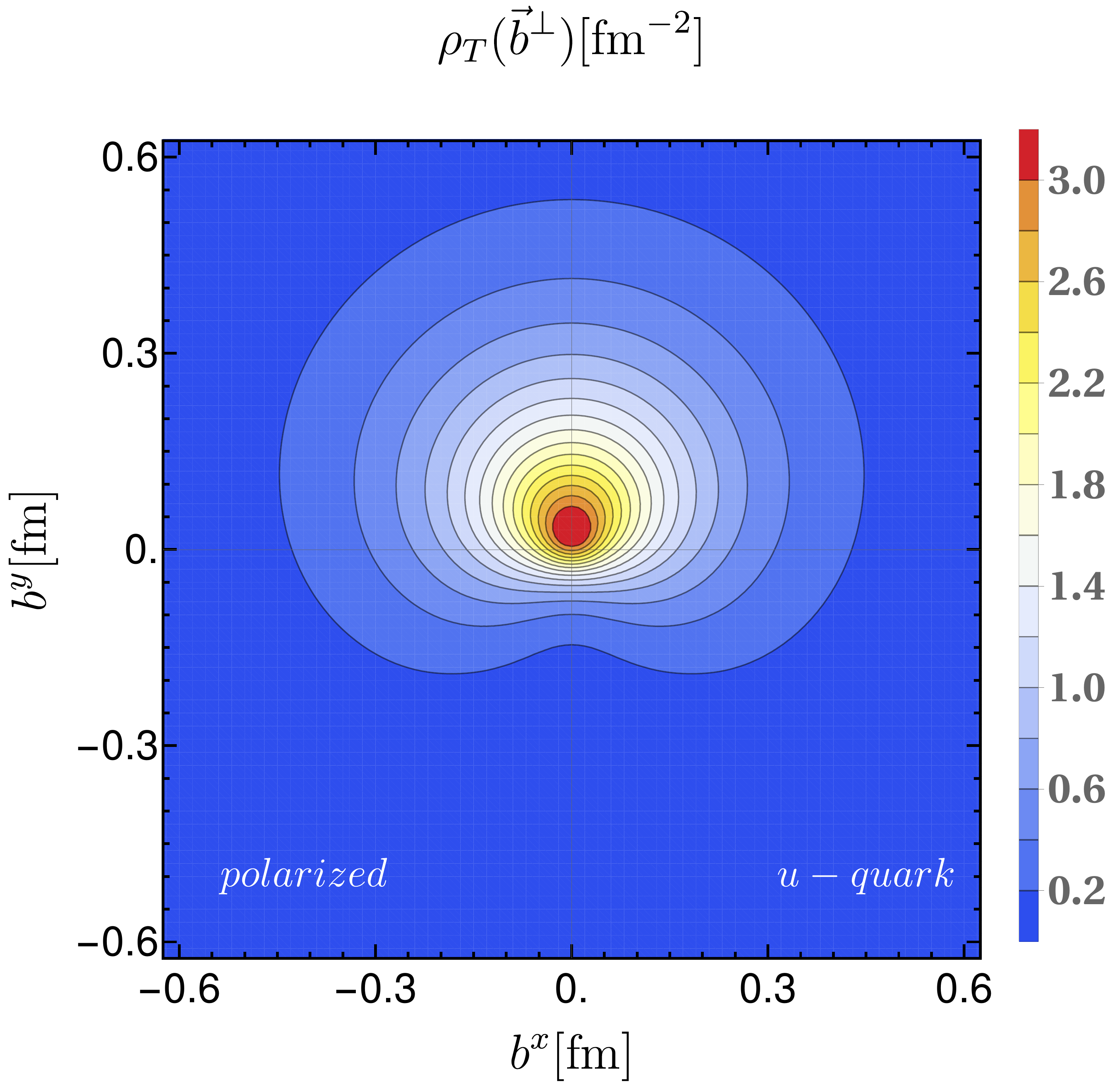}}
\end{tabular}
\begin{tabular}{cc}
\subfloat[]{\includegraphics[scale=0.25]{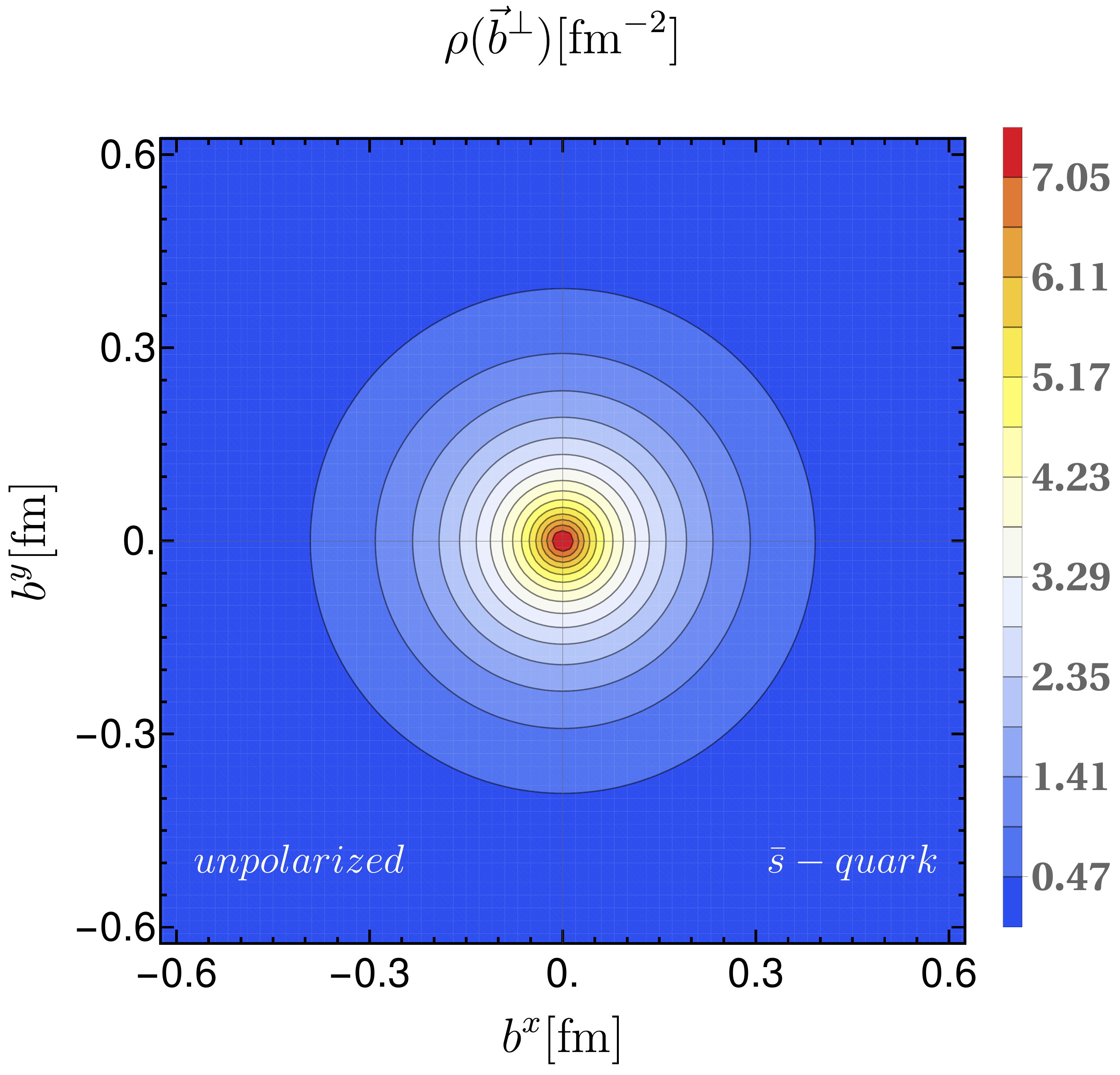}}
\end{tabular}
\begin{tabular}{cc}
\subfloat[]{\includegraphics[scale=0.25]{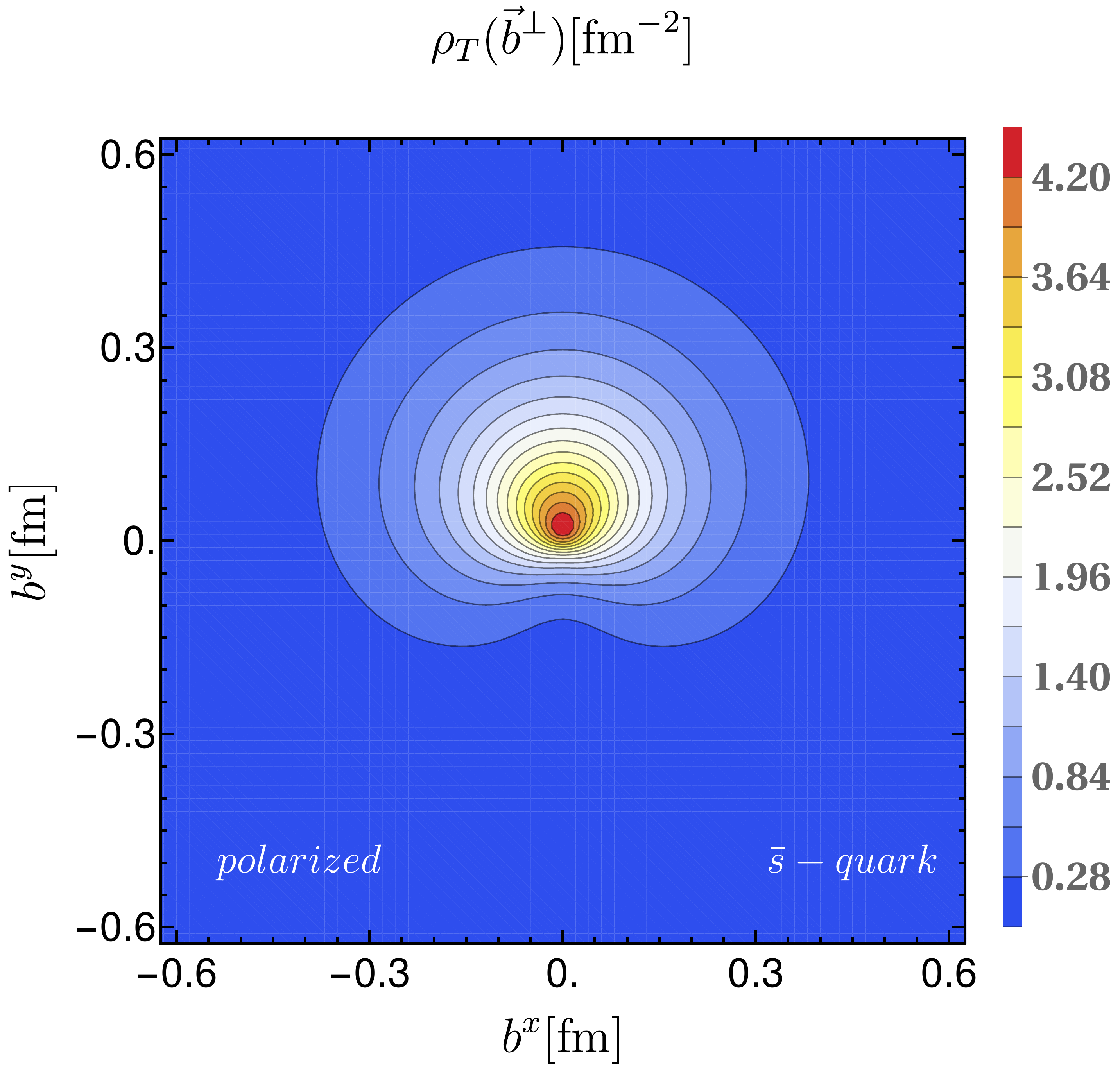}}
\end{tabular}
\caption{The valence quarks' probability densities in the kaon in the transverse impact parameter plane; (a) for unpolarized $u$ quark $\rho({\vec b}^{\perp})=\rho^{n=1}({\vec b}^{\perp},{\vec s}^{\perp}=\vec{0})$ and (b) for polarized $u$ quark  $\rho_T({\vec b}^{\perp})=\rho^{n=1}({\vec b}^{\perp},{\vec s}^{\perp}=(1,0))$. (c) and (d) are same as (a) and (b), respectively but for $\bar{s}$ quark in the kaon.}
\label{kaon_density}
\end{figure*}
\begin{figure*}
\begin{tabular}{cc}
\subfloat[]{\includegraphics[scale=0.3]{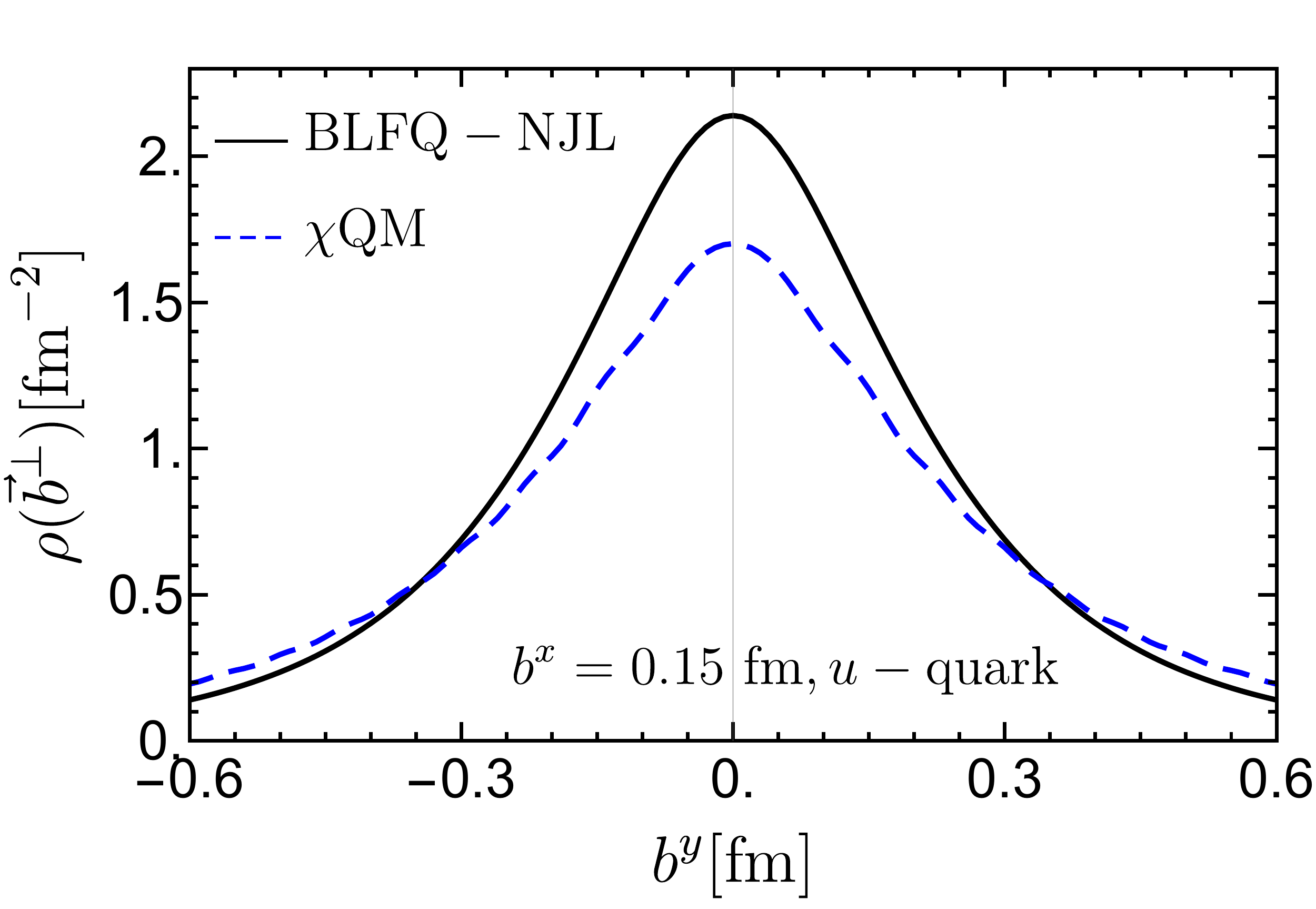}}
\end{tabular}
\begin{tabular}{cc}
\subfloat[]{\includegraphics[scale=0.3]{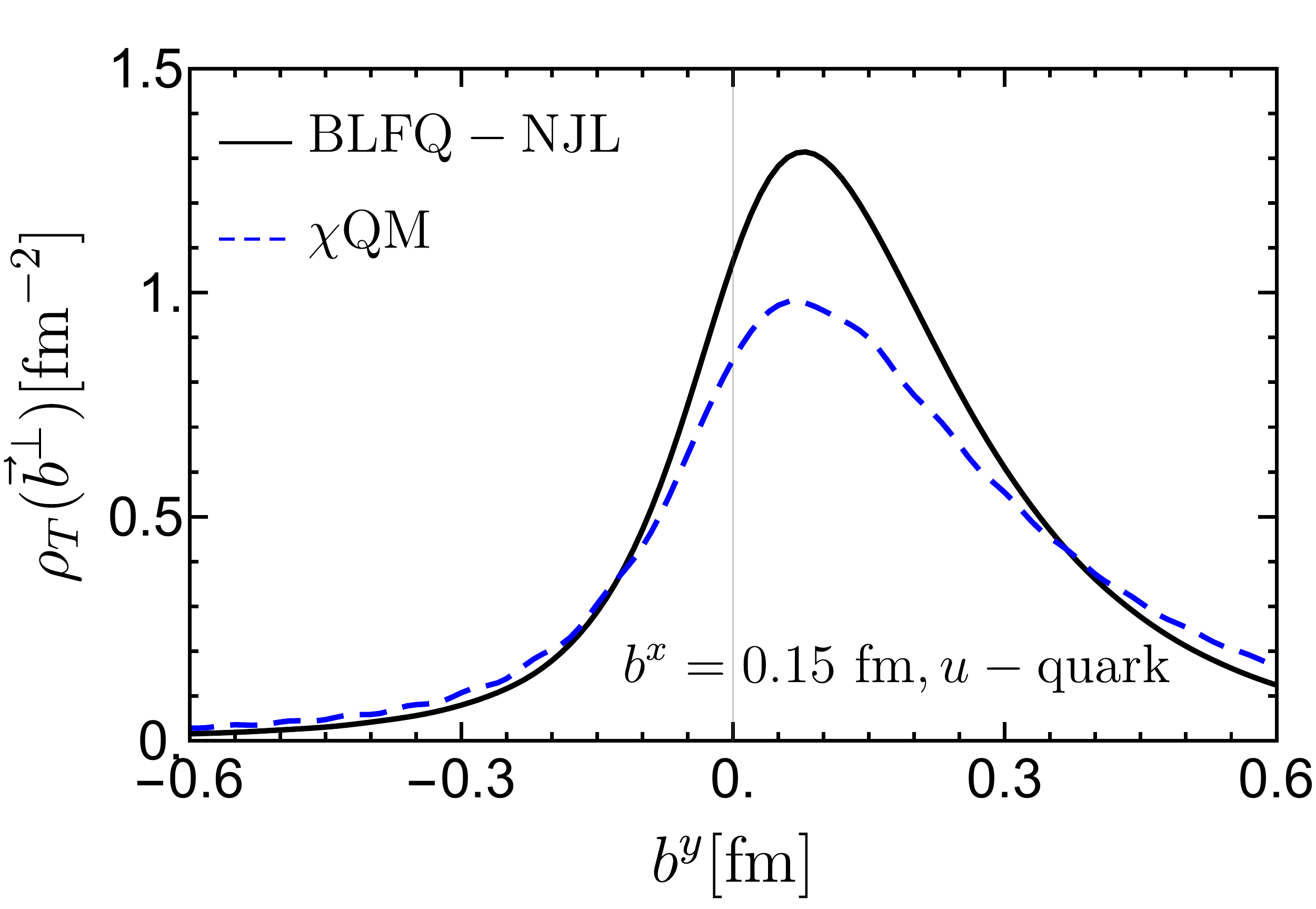}}
\end{tabular}
\begin{tabular}{cc}
\subfloat[]{\includegraphics[scale=0.3]{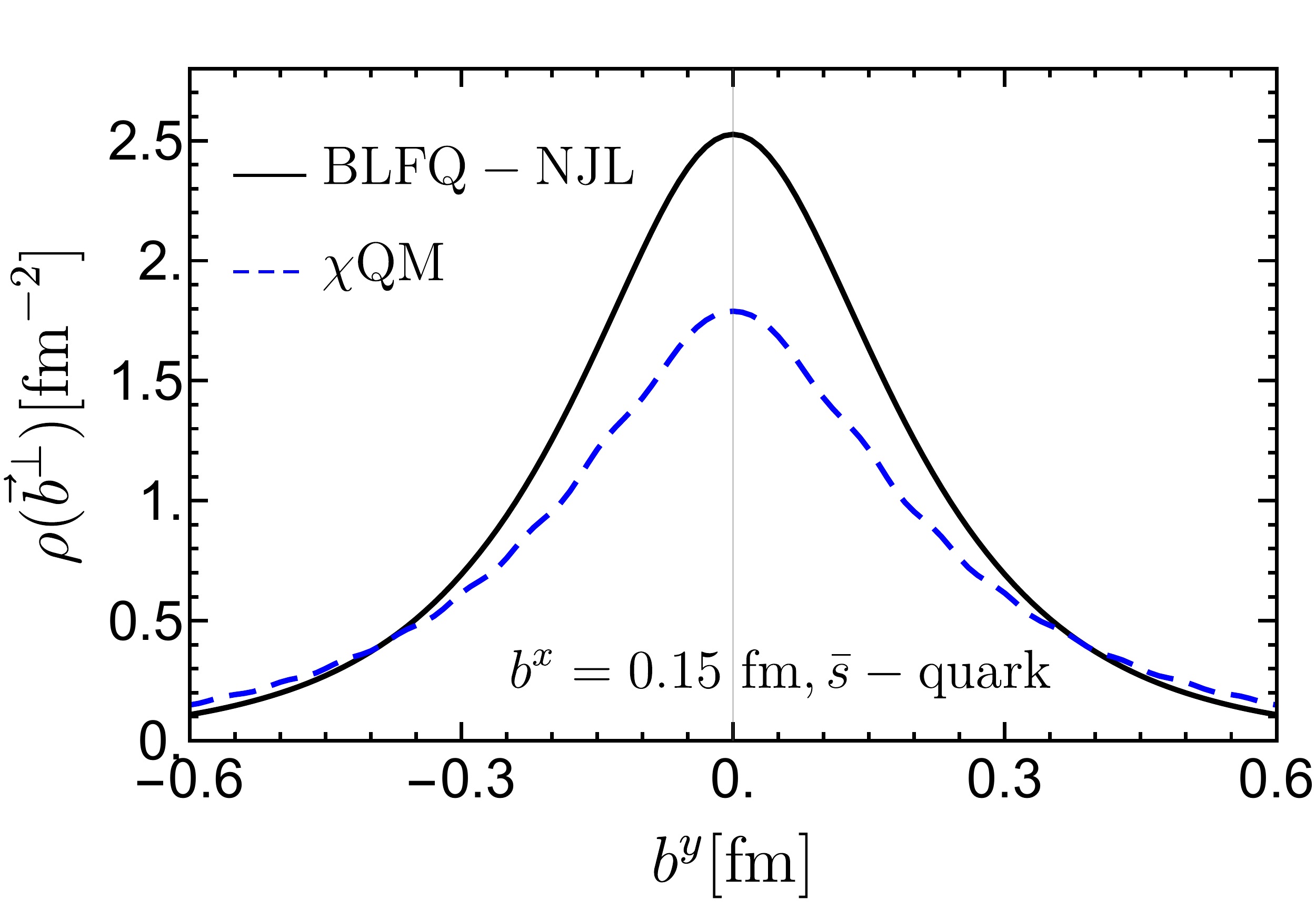}}
\end{tabular}
\begin{tabular}{cc}
\subfloat[]{\includegraphics[scale=0.3]{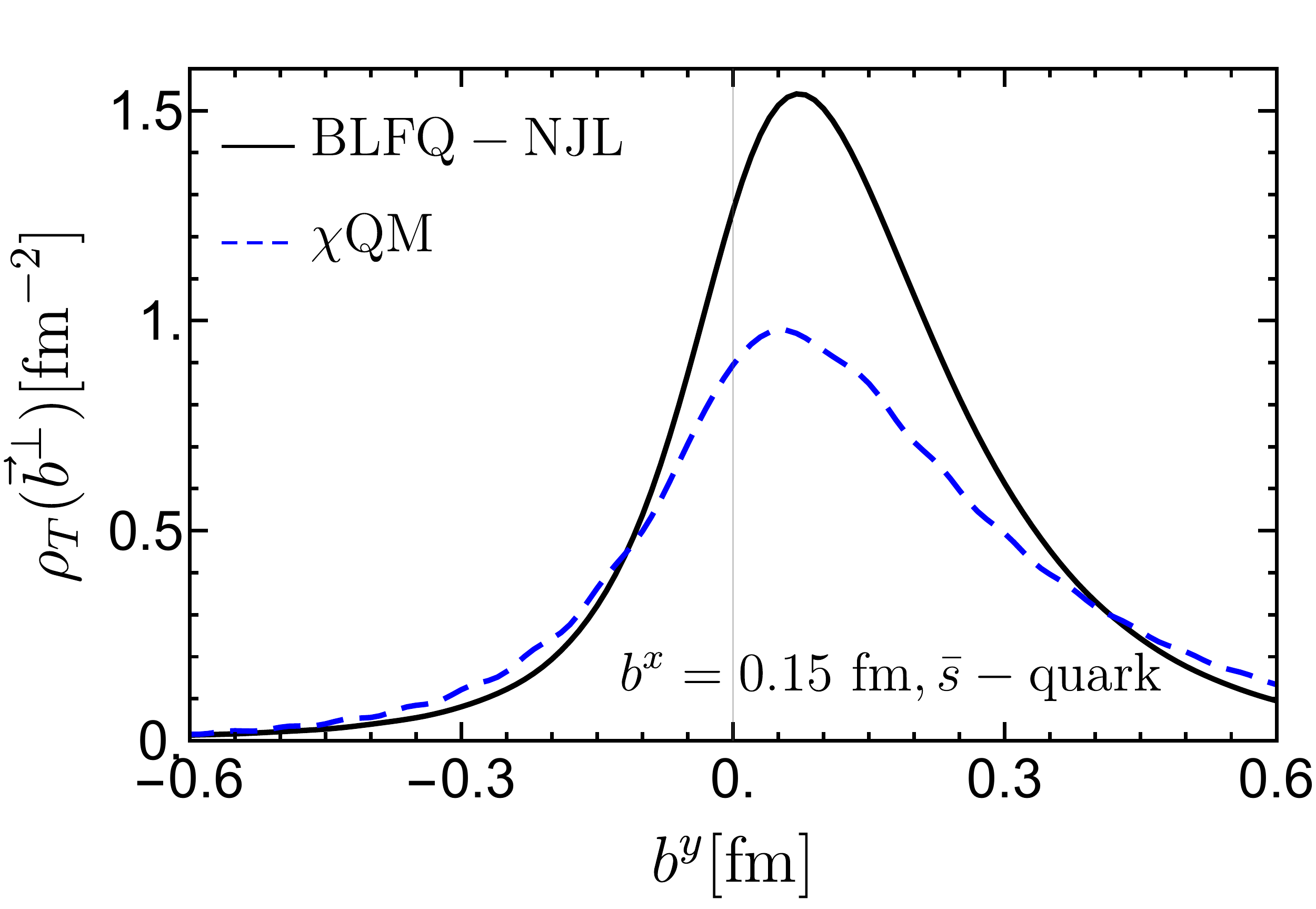}}
\end{tabular}
\caption{The valence quarks' probability densities in the kaon in our BLFQ-NJL model (solid lines) are compared with $\chi$QM predictions~\cite{Nam:2011yw} (dashed lines).}
\label{kaon_density_2d}
\end{figure*}
\subsection{Spin densities of the pion and the kaon \label{densities}}
The GPDs in the transverse impact parameter space at  zero skewness can be interpreted as the densities of quarks with longitudinal momentum fraction $x$ and transverse location ${\vec b}^{\perp}$ with respect to the center of momentum of the hadron independent of the polarization. On the one hand, the density $\rho(x,{\vec b}^{\perp},\lambda)$ of quarks with helicity $\lambda$ 
in the pion (kaon) is determined by the unpolarized density, $2\rho(x,{\vec b}^{\perp},\lambda) = q(x,{\vec b}^{\perp})$, where the latter is the ${\vec b}^{\perp}$-dependent GPD at zero skewness given by Eq.~(\ref{eq:Hb}). On the other hand, the density of quarks with transverse spin $\vec{s}^\perp$, $\rho(x,{\vec b}^{\perp},{\vec s}^{\perp})$, in the pion (kaon) can be expressed in a combination of the GPDs $q(x,{\vec b}^{\perp})$ and $q_T(x,{\vec b}^{\perp})$ as
\begin{align}
\rho(x,{\vec b}^{\perp},{\vec s}^{\perp})=\frac{1}{2} \left[ q(x,{\vec b}^{\perp}) 
  - \frac{{\vec s}^{\perp}_i \epsilon_{ij}^{\perp} {\vec b}^{\perp}_j}{M_{\mathcal{P}}}\, 
q_T^{\prime}(x,{\vec b}^{\perp})
  \,\right] \,,
\end{align}
where $q_T^{\prime}(x,{\vec b}^{\perp})=\frac{\partial}{\partial (b^\perp)^2}\,q_T(x,{\vec b}^{\perp})$. The quark spin densities have been investigated in Refs.~\cite{Diehl:2005jf,Gockeler:2006zu,Pasquini:2007xz,Maji:2017ill} for quarks with transverse spin ${\vec s}^{\perp}$ in the nucleon having transverse spin~(${\vec S}^{\perp}$).  The corresponding expression for transversely polarized quarks in the pseudoscalar mesons is achieved by setting ${\vec S}^{\perp}=0$ in the nucleon densities~\cite{Gockeler:2006zu}. One finds that the result is  much simpler but still involves a dipole term $\propto {\vec s}^{\perp}_i \epsilon_{ij}^{\perp} {\vec b}^{\perp}_j$ leading to a 
dependence on the direction of ${\vec b}^{\perp}$ for fixed ${\vec s}^{\perp}$. The $x$-moments of quark spin densities are then given by~\cite{Brommel:2007xd}
\begin{align}
 \rho^{n}({\vec b}^{\perp},{\vec s}^{\perp})
 &= \int_{0}^{1} dx\, x^{n-1} \rho(x,{\vec b}^{\perp},{\vec s}^{\perp}) \nonumber\\&
= \frac{1}{2} \left[ \mathcal{A}^q_{n0}({\vec b}^{\perp}) 
  - \frac{{\vec s}^{\perp}_i \epsilon_{ij}^{\perp} {\vec b}^{\perp}_j}{M_{\mathcal{P}}}\,
\mathcal{B}_{Tn0}^{q\prime}({\vec b}^{\perp})
  \,\right] \,,
\label{density}
\end{align}
where the ${\vec b}^{\perp}$-dependent vector and tensor generalized FFs, $\mathcal{A}^q_{n0}$ and $\mathcal{B}^q_{Tn0}$, are obtained  by performing the Fourier transform of the FFs $A^q_{n0}(t)$ and $B^q_{Tn0}(t)$ with respect to $\vec{\Delta}^\perp$ or equivalently by taking the $x$ moments of the impact parameter dependent GPDs $q(x,{\vec b}^{\perp})$ and $q_T(x,{\vec b}^{\perp})$, respectively:
\begin{align}
\mathcal{A}^q_{n0}({\vec b}^{\perp})&=\int \frac{d^2{\vec \Delta}^\perp}{(2\pi)^2}
e^{-i {\vec \Delta}^\perp \cdot {\vec b}^\perp } {A}^q_{n0}(-{\vec \Delta}^{\perp 2})\nonumber\\&=\int_{0}^1 dx\, x^{n-1} q(x,{\vec b}^{\perp})  \,,
\nonumber\\
\mathcal{B}^q_{T n0}({\vec b}^{\perp})&=\int \frac{d^2{\vec \Delta}^\perp}{(2\pi)^2}
e^{-i {\vec \Delta}^\perp \cdot {\vec b}^\perp } {B}^q_{Tn0}(-{\vec \Delta}^{\perp 2})\nonumber\\&=\int_{0}^1 dx\, x^{n-1} q_T(x,{\vec b}^{\perp})  \,.
  \label{EqGPDmoments}
\end{align}

The impact parameter dependent GPDs $q(x,{\vec b}^{\perp})$ and $q_T(x,{\vec b}^{\perp})$ for the pion are presented in Fig.~\ref{pion_impact_gpds}. We find that both distributions have sharp peaks located at the center of the pion ($b^\perp=0$) when the quark carries large longitudinal momentum. Nevertheless, the magnitude of the unpolarized distribution is much higher compared
to that of the polarized distribution. A substantial difference is also observed between $q(x,{\vec b}^{\perp})$ and $q_T(x,{\vec b}^{\perp})$ at large $x$. We also notice that the qualitative behavior of the GPDs $q(x,{\vec b}^{\perp})$ and $q_T(x,{\vec b}^{\perp})$ for the kaon, shown in Fig.~\ref{kaon_impact_gpds}, is very similar to those for the pion. However, due to the heavier mass of the $\bar{s}$ quark, its distributions are narrower than those distributions for the $u$ quark in the kaon. Another interesting feature is that the width of all the GPDs in the transverse impact parameter space decrease as $x$ increases. This indicates that the distributions are more localized near the center of the momentum ($b_\perp=0$) when quarks are carrying higher longitudinal momentum. This characteristic of the GPDs in the transverse impact parameter space is reassuring since the distributions in the momentum space become broader in $-t$ with increasing $x$, as can be seen from Figs.~\ref{pion_gpds} and \ref{kaon_gpds}.  On the light-front, this is understood as the larger the momentum fraction, the lower the kinetic energy carried by the quarks. As the total kinetic energy remains limited, the distribution in the transverse momentum is required to become broader to carry a larger portion of the kinetic energy. This model-independent property of the GPDs is also observed in the case of the nucleon~\cite{Chakrabarti:2013gra,Mondal:2015uha,Chakrabarti:2015ama,Maji:2017ill}.

We present the first moment of the quark-spin probability density $\rho^{n=1}({\vec b}^{\perp},{\vec s}^{\perp})$, defined in Eq.~(\ref{density}), in Fig.~\ref{pion_density}. When the quark is unpolarized (${\vec s}^{\perp}= \vec{0}$), only $\mathcal{A}^q_{10}({\vec b}^{\perp})$ contributes in the probability density, which is rotationally symmetric in the two-dimensional impact parameters $(b_x,b_y)$ plane as shown in Fig.~\ref{pion_density}(a) and hence, one does not see any interesting structures from this. We now turn our attention to the case when the quark is transversely polarized. Without loss of generality,  we consider the quark polarized along the $x$-axis, i.e., ${\vec s}^{\perp}= (+1,0)$ and show
the numerical results as functions of $b_x$ and $b_y$. The probability density becomes distorted when the quark inside the pion is transversely polarized as can be seen from Fig.~\ref{pion_density}(b) indicating the spin structure inside the pion. The second
term in Eq.~(\ref{density}) provides the distortion, and one can clearly observe the deviation from rotational symmetry of the unpolarized density due to the polarization. We also find that the present results are very similar to those given by the lattice QCD
calculation~\cite{Brommel:2007xd}. For instance, in the lower panel of Fig.~\ref{pion_density} we show the probability densities as a function of $b_y$ at fixed $b_x=0.15$ fm, comparing those with that of the lattice QCD simulations and the $\chi$QM~\cite{Nam:2010pt}. The BLFQ-NJL model results are found to be consistent with the results of lattice QCD and the $\chi$QM. The spin densities of the $u$ and $\bar{s}$ quarks in the kaon are shown in Fig.~\ref{kaon_density}, where we notice the similar patterns of the quark-spin probability densities as observed in the pion. It is however interesting to note that $\bar{s}$ quark densities, due to the heavier $\bar{s}$  mass, are more localized near the origin compared to the $u$-quark densities in the kaon. We also find that the qualitative behavior of the present results in the BLFQ-NJL model is compatible with the results obtained in the $\chi$QM~\cite{Nam:2011yw} as shown in Fig.~\ref{kaon_density_2d}, where we plot the probability densities as a function of $b_y$ at fixed $b_x=0.15$ fm. 
\subsection{Average transverse shift and transverse squared radius \label{transverse_shift}}
It is also interesting to examine the average transverse shift of the peak position of the probability density along the $b_y$ direction for a transverse quark spin in the $x$-direction, which is defined as~\cite{Brommel:2007xd}
\begin{align}
  \label{shift}
\langle b_y^\perp \rangle_{n}
 &= \frac{\int d^2 \vec{b}^\perp\, b_{y}^\perp\, \rho^{n}({\vec b}^{\perp},{\vec s}^{\perp})}{%
          \int d^2 \vec{b}^\perp\, \rho^{n}({\vec b}^{\perp},{\vec s}^{\perp})}
= \frac{1}{2 M_{\mathcal{P}}}\, \frac{B^q_{T n0}(0)}{A^{q}_{n0}(0)}.
\end{align}
Our BLFQ-NJL model results for the pion give $\langle b_y^\perp \rangle_1=0.162\pm 0.003$ fm and $\langle b_y^\perp \rangle_2=0.131\pm 0.003$ fm, while the lattice simulations provide~\cite{Brommel:2007xd} $\langle b_y^\perp \rangle_1 = 0.151(24)$ fm  and $\langle b_y^\perp \rangle_2 = 0.106(28)$ fm. Our results for $\langle b_y^\perp \rangle_{n=1,2}$ for the pion and the kaon are compared with the lattice QCD, the $\chi$QM, CCQM, and NJL model in Table~\ref{tab:shift}.
 \begin{table*}
\caption{BLFQ-NJL model predictions for average transverse shift $\langle b_y^\perp \rangle_{1,2}^{q}$ in the pion and the kaon. Our results are compared with the available lattice simulations~\cite{Brommel:2007xd}, the $\chi$QM~\cite{Nam:2010pt,Nam:2011yw}, CCQM~\cite{Fanelli:2016aqc}, and NJL model~\cite{Zhang:2021tnr}.
}\label{tab:shift}
 \centering
\begin{tabular}{ccc ccc ccc ccc ccc ccc ccc}
\toprule
 Approach ~&~ $\langle b_y^\perp \rangle_{1}^{q,\pi}$ fm ~&~ $\langle b_y^\perp \rangle_{2}^{q,\pi}$ fm ~&~ $\langle b_y^\perp \rangle_{1}^{u,K}$ fm ~&~ $\langle b_y^\perp \rangle_{1}^{\bar{s},K}$ fm ~&~ $\langle b_y^\perp \rangle_{2}^{u,K}$ fm ~&~ $\langle b_y^\perp \rangle_{2}^{\bar{s},K}$ fm &\\

\colrule

BLFQ-NJL (this work) & $0.162\pm 0.003$ & $0.131\pm 0.003$  & $0.164\pm 0.003$  & $0.141\pm 0.002$  & $0.114\pm 0.002$  & $0.114\pm 0.002$ &\\
\vspace{0.1cm}
Latiice QCD~\cite{Brommel:2007xd} & 0.151 $\pm$ 0.024 & 0.106 $\pm$ 0.028 & ...  & ...  & ...  & ...  & \\
\vspace{0.1cm}
$\chi$QM~\cite{Nam:2010pt} & 0.152 & ...  & ... & ... & ... & ... & \\
\vspace{0.1cm}
$\chi$QM (model I)~\cite{Nam:2011yw} & ...  & ...  & 0.168  & 0.166 & ... & ... &\\
\vspace{0.1cm}
$\chi$QM (model II)~\cite{Nam:2011yw} & ...  & ...  & 0.139  & 0.100 & ... & ... &\\
\vspace{0.1cm}
CCQM~\cite{Fanelli:2016aqc} & $0.090\pm 0.001$  & $0.080\pm 0.001$  & ...  & ... & ... & ... &\\
NJL model~\cite{Zhang:2021tnr} & ...  & ...  & 0.116  & ... & 0.083 & ... &\\
\botrule
\end{tabular}
\end{table*}
\begin{figure*}[htbp]
\centering
\includegraphics[scale=0.35]{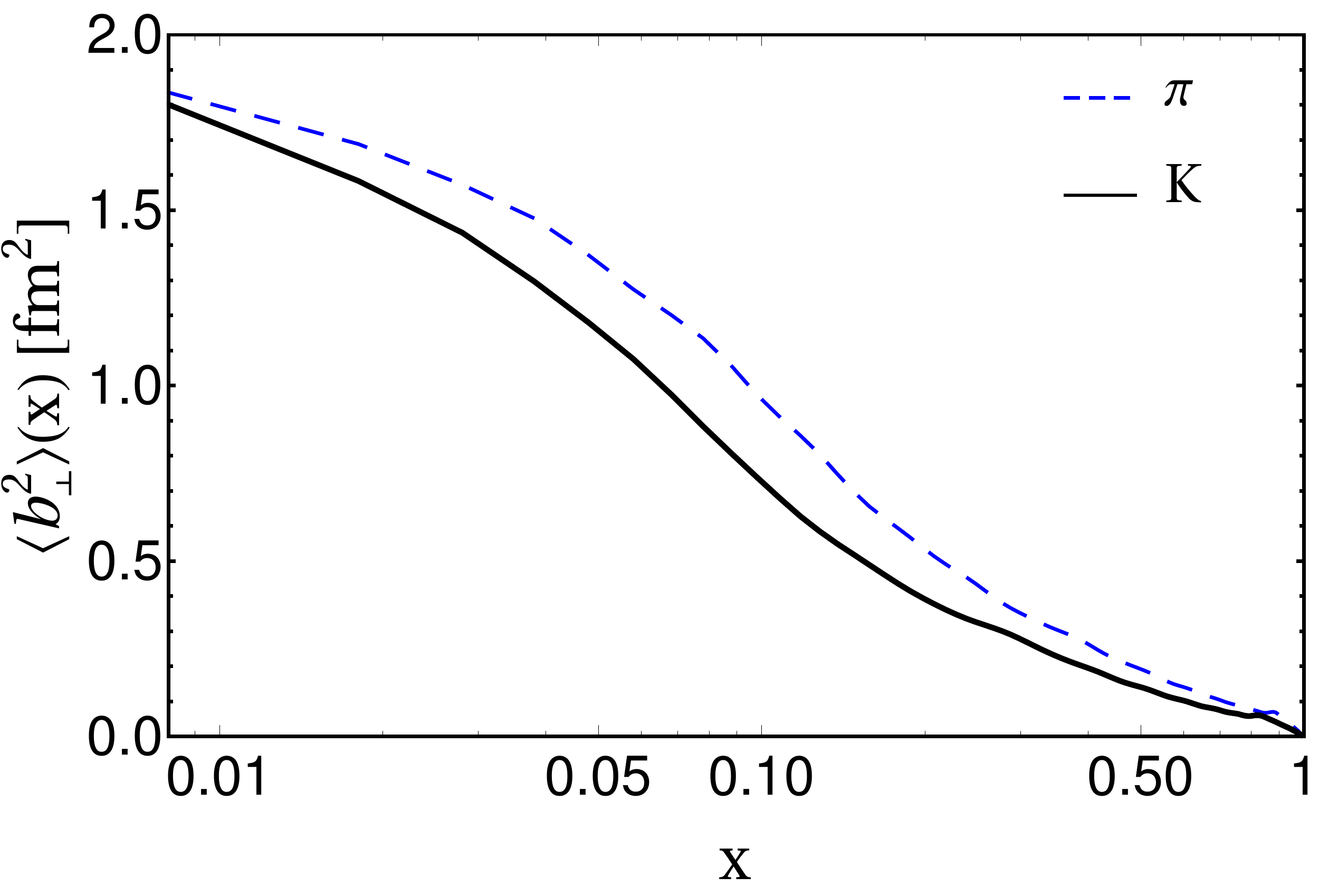}
\caption{$x$-dependence of  $\langle b^2_\perp \rangle$ for quarks in the pion (dashed line) and the kaon (solid line) in the BLFQ-NJL model.}
\label{b2x}
\end{figure*}

One can also define the $x$-dependent squared radius of the quark density in the transverse plane as~\cite{Dupre:2016mai}:
\begin{align}
\langle b_{\perp}^2 \rangle^q (x) = \frac{ \int d^2 {\vec{ b}^\perp} \, ({\vec b}^{\perp})^2 q(x, {\vec b}^{\perp})}{\int d^2 {\vec{ b}^\perp}\,  q(x, {\vec b}^{\perp})},
\label{eq:b2}
\end{align}
which can also be written through the GPD $H(x,0,t)$ as:
\begin{equation}
\langle b_{\perp}^2 \rangle^q (x)= 4 \frac{\partial}{\partial t} \ln H^q(x,0,t) \biggr| _{t = 0}.
\label{eq:crgpd}
\end{equation}
For the pion and the kaon the squared radius $\langle b_{\perp}^2 \rangle(x)$ is obtained as the charge-weighted sum over the valence quarks: $\langle b_{\perp}^2 \rangle^\pi(x)=e_u \langle b_{\perp}^2 \rangle^u(x) + e_{\bar{d}}\langle b_{\perp}^2 \rangle^{\bar{d}}(x)$ and $\langle b_{\perp}^2 \rangle^K(x)=e_u \langle b_{\perp}^2 \rangle^u(x) + e_{\bar{s}}\langle b_{\perp}^2 \rangle^{\bar{s}}(x)$. The
$\langle b^2_\perp \rangle (x)$ describes the transverse
size of the hadron and shows an increase of transverse radius with decreasing value of the quark momentum fraction $x$~\cite{Dupre:2016mai}. As can be seen from Fig.~\ref{b2x} and as expected, the transverse size of the kaon is smaller than that of the pion for a fixed value of $x$. We also compute the pion's and the kaon's transverse squared radius through the following average over $x$~\cite{Dupre:2016mai}
\begin{equation}
\langle b_{\perp}^2 \rangle = \sum_q e_q \frac{1}{N_q}\int_0^1 dx H^q(x,0,0) {\langle b_{\perp}^2 \rangle}^q(x)\,,
\end{equation}
with the integrated number of valence quark $N_q$ of flavor $q$.
We obtain the squared radius of the pion and the kaon, $\langle b_{\perp}^2 \rangle^\pi=0.285$ fm$^2$ and $\langle b_{\perp}^2 \rangle^K=0.223$ fm$^2$, respectively. The quantity $\langle b_{\perp}^2 \rangle$ is connected to the conventionally defined squared radius $\langle r_{c}^2 \rangle$ from the EMFF by $\langle b_{\perp}^2 \rangle=\frac{2}{3} \langle r_{c}^2 \rangle$~\cite{Dupre:2016mai,Li:2017mlw}. Our results are close to the experimental data for the pion, $\langle b_{\perp}^2 \rangle^\pi_{\rm exp}=0.301\pm 0.014$ fm$^2$ and for the kaon, $\langle b_{\perp}^2 \rangle^K_{\rm exp}=0.209\pm 0.047$ fm$^2$~\cite{ParticleDataGroup:2018ovx} and also consistent with the previously computed charge radii of the pion and the kaon in the BLFQ-NJL model~\cite{Jia:2018ary}.

\section{Summary \label{summary}}
We have investigated the valence quark GPDs of the light pseudoscalar mesons in the framework of BLFQ using a light-front model for light mesons that incorporates light-front holography, longitudinal confinement, and the color-singlet Nambu-Jona--Lasinio interactions. The parameters in the BLFQ-NJL model have previously  been adjusted to generate the experimental mass spectrum and the charge radii of the light mesons~\cite{Jia:2018ary}. We have evaluated the quark unpolarized GPD $H$ and tensor GPD $E_T$ in the pion and the kaon in both momentum and transverse position space. The generalized form factors for the pion and the kaon, i.e., vector and tensor form factors from the first two moments of the quark unpolarized and tensor  GPDs have been calculated. We have verified the agreement of the the electromagnetic form factors resulting from the unpolarized GPD with the experimental data of the pion and the kaon. The moments of the tensor GPD $E_T$, which give the tensor form factors, have been found to be comparable with the  parameterization of lattice QCD simulations as well as with the results of the  $\chi$QM. 

We have subsequently calculated  the probability densities of the unpolarized and polarized quarks inside the pion and the kaon. We have observed that the spatial distribution of the unpolarized quarks is axially symmetric, while it strongly distorted when quarks are transversely polarized, revealing a non-trivial distribution of quark polarization in the pseudoscalar mesons. 
The  quark probability densities in the BLFQ-NJL model have been found to be in good agreement to those from lattice QCD. The qualitative nature of the quark densities in the kaon was also consistent with those in the $\chi$QM.
In order to examine the shift of the peaks of the densities in the $b_y$ direction, we have computed the average value of $b_y$,  which turned out to be compatible with the lattice QCD and the $\chi$QM.

We have also evaluated the $x$-dependent squared radius of the
quark density in the transverse plane, which describes the transverse size of the hadron. We have found that, with increasing quark longitudinal momentum, the transverse radius of the pion and the kaon decreases. A similar effect has also been observed in the nucleon~\cite{Dupre:2016mai}. We have noticed that the quarks are more transversely localized in the kaon than in the pion.

\section*{ACKNOWLEDGMENTS}
C. M. is supported by new faculty start up funding by the Institute of Modern Physics, Chinese Academy of Sciences, Grant No. E129952YR0. 
C. M. and S. N. thank the Chinese Academy of Sciences Presidents International Fellowship Initiative for the support via Grants No. 2021PM0023 and 2021PM0021, respectively.  X. Z. is supported by new faculty startup funding by the Institute of Modern Physics, Chinese Academy of Sciences, by Key Research Program of Frontier Sciences, Chinese Academy of Sciences, Grant No. ZDB-SLY-7020, by the Natural Science Foundation of Gansu Province, China, Grant No. 20JR10RA067 and by the Strategic Priority Research Program of the Chinese Academy of Sciences, Grant No. XDB34000000. J. P. V. is supported by the Department of Energy under Grants No. DE-FG02-87ER40371, and No. DE-SC0018223 (SciDAC4/NUCLEI). S. J. is  supported by U.S. Department of Energy, Office of Science, Office of Nuclear Physics, contract no. DE-AC02-06CH11357. This research used resources of the National Energy Research Scientific Computing Center (NERSC), a U.S. Department of Energy Office of Science User Facility operated under Contract No. DE-AC02-05CH11231. A portion of the computational resources were also provided by Gansu Computing Center.

\end{document}